\documentclass[reprint,amsmath,amssymb,aps,]{revtex4-2}

\usepackage{graphicx}	
\usepackage{dcolumn}	
\usepackage{bm}			
\usepackage{hyperref}
\usepackage{amsmath}
\usepackage{verbatim}
\usepackage{slashed}
\usepackage{hyperref}
\usepackage{mathtools}
\usepackage{tabularx}
\hypersetup{
    colorlinks,
    citecolor=blue,
    filecolor=black,
    linkcolor=blue,
    urlcolor=black
}
\begin{document}

\title{A Flavorful Composite Higgs Model : \\Connecting the B anomalies with the hierarchy problem}
\author{Yi Chung}
\email{yichung@ucdavis.edu}
\affiliation{
 Center for Quantum Mathematics and Physics (QMAP), Department of Physics, \\University of California,  Davis, CA 95616, U.S.A.
}

\begin{abstract}
We present a model which connects the neutral current B anomalies with composite Higgs models. The model is based on the minimal fundamental composite Higgs model with $SU(4)/Sp(4)$ coset. The strong dynamics spontaneously break the symmetry and introduce five Nambu-Goldstone bosons. Four of them become the Standard Model Higgs doublet and the last one, corresponding to the broken local $U(1)'$ symmetry, is eaten by the gauge boson. This leads to an additional TeV-scale $Z'$ boson, which can explain the recent B anomalies. The experimental constraints and allowed parameter space are discussed in detail.
\end{abstract}

\maketitle

\section{Introduction}


The Standard Model (SM) of particle physics successfully describes all known elementary particles and their interactions. However, there are still a few puzzles that have yet to be understood. One of them is the well-known hierarchy problem. With the discovery of light Higgs bosons in 2012 \cite{Chatrchyan:2012xdj, Aad:2012tfa}, the last missing piece of the SM seemed to be filled. However, SM does not address the UV-sensitive nature of scalar bosons. The Higgs mass-squared receives quadratically divergent radiative corrections from the interactions with SM fields, which require an extremely sensitive cancellation to get a $125$ GeV Higgs boson. To avoid the large quadratic corrections, the most natural way is to invoke some new symmetry such that the quadratic contributions cancel in the symmetric limit. This requires the presence of new particles related to SM particles by the new symmetry.


One appealing solution to the hierarchy problem is the composite Higgs model (CHM), where the Higgs doublet is the pseudo-Nambu-Goldstone boson (pNGB) of a spontaneously broken global symmetry of the underlying strong dynamics~\cite{Kaplan:1983fs, Kaplan:1983sm}. Through the analogy to the chiral symmetry breaking in quantum chromodynamics (QCD), which naturally introduces light scalar fields, i.e., pions, we can construct models with light Higgs bosons in a similar way. In a CHM, an approximate global symmetry $G$ is spontaneously broken by some strong dynamics down to a subgroup $H$ at a symmetry breaking scale $f$. The heavy resonances of the strong dynamics are expected to be around the compositeness scale $\sim 4\pi f$ generically. The pNGBs of the symmetry breaking, on the other hand, can naturally be light with masses $<f$ as they are protected by the shift symmetry.

Among all types of CHMs with different cosets, the CHMs with fundamental gauge dynamics featuring only fermionic matter fields are of interest in many studies \cite{Barnard:2013zea, Ferretti:2013kya, Cacciapaglia:2014uja, Cacciapaglia:2020kgq}, which is known as the fundamental composite Higgs model (FCHM). In this type of CHMs, hyperfermions $\psi$ are introduced as the representation of hypercolor (HC) group $G_{HC}$. Once the HC group becomes strongly coupled, hyperfermions form a condensate, which breaks the global symmetry. However, they always introduce more than four pNGBs, which means more light states are expected to be found. The minimal FCHM, which is based on the $SU(4)/Sp(4)$ coset \cite{Katz:2005au, Gripaios:2009pe, Galloway:2010bp}, contains five pNGBs. The four of them formes the SM Higgs doublet, and the fifth one, as a SM singlet, could be a light scalar boson (if the symmetry is global) or a TeV-scale $Z'$ boson (if the symmetry is local). No matter which, it should lead to some deviations in low energy phenomenology.


Although the direct searches by ATLAS and CMS haven't got any evidence of new particles, LHCb, which does the precise measurement of B meson properties, shows interesting hints of new physics. There are discrepancies in several measurements of semileptonic B meson decays, especially the tests of lepton flavor universality (LFU), which are so-called the neutral current B anomalies \cite{LHCb:2013ghj, LHCb:2014vgu, LHCb:2015svh, LHCb:2017avl, LHCb:2019hip, LHCb:2020lmf, LHCb:2021trn}. Each anomaly is not statistically significant enough to reach the discovery level, but the combined analysis shows a consistent deviation from the SM prediction \cite{Altmannshofer:2021qrr, Cornella:2021sby, Geng:2021nhg, Alok:2019ufo, Alguero:2021anc, Carvunis:2021jga}. These anomalies might be the deviation we are looking for.


One of the popular explanations is through a new $Z'$ vector boson which has flavor-dependent interactions with SM fermions. Many different types of $Z'$ models with diverse origins of $U(1)'$ gauge symmetry have been proposed 
\cite{Altmannshofer:2014cfa, Altmannshofer:2015mqa, Altmannshofer:2019xda, Crivellin:2015mga, Crivellin:2016ejn, Alonso:2017bff, Alonso:2017uky, Bonilla:2017lsq, Allanach:2020kss, Allanach:2018lvl, Allanach:2019iiy, Gauld:2013qba, Buras:2013dea, Buras:2013qja, AristizabalSierra:2015vqb, Celis:2015ara, Falkowski:2015zwa, Chiang:2016qov, Boucenna:2016wpr, Boucenna:2016qad, Bhatia:2017tgo, Ko:2017lzd, Tang:2017gkz, Fuyuto:2017sys, Bian:2017xzg, King:2018fcg, Duan:2018akc, Calibbi:2019lvs}. 
Depending on its couplings with fermions, the mass of the $Z'$ can range from sub-TeV to multi-TeV. For a $Z'$ boson at the TeV scale, it is natural to try to connect it with the hierarchy problem \footnote{For our interest, we would like to mention some researches aiming at explaining the B anomalies within composite Higgs models. Different studies using different features of composite theory to address the problem, such as additional composite leptoquarks \cite{Gripaios:2014tna, Barbieri:2016las, Marzocca:2018wcf, Fuentes-Martin:2020bnh} or composite vector resonances \cite{Niehoff:2015bfa, Niehoff:2015iaa, Carmona:2015ena, Carmona:2017fsn, Barbieri:2017tuq, Sannino:2017utc, Chala:2018igk}. However, they are all different from this study, where we introduce a new $Z'$ boson.}.


In this paper, we realize this idea using a $SU(4)/Sp(4)$ FCHM, where an $U(1)'$ subgroup within $SU(4)$ is gauged. The corresponding $Z'$ boson only couples to the third generation SM fermions $F_3$ and the hyperfermions $\psi$ through the terms
\begin{align}
\mathcal{L}_{\text{int}}=g_{Z'}Z'_\mu\,(\,\bar{F}_3\gamma^\mu F_3 + Q_{HC}\bar{\psi}_{}\gamma^\mu \psi_{}\,),
\label{int}
\end{align}
where $g_{Z'}$ was normalized such that SM fermions $F_3$ carry a unit charge and hyperfermions carry charge $Q_{HC}$. When the hypercolor group becomes strongly coupled, the global symmetry $SU(4)$ and its gauged $U(1)'$ subgroup are broken. The $5^{th}$ pNGB is eaten by the $U(1)'$ gauge boson, which results in a TeV-scale $Z'$ boson. We will test the potential for this $Z'$ boson to explain the neutral current B anomalies. The parameter space allowed by different experimental constraints, mainly from neutral meson mixings and lepton flavor violation decays, will be discussed. The bounds on $M_{Z'}$ from the LHC direct searches are also shown.


This paper is organized as follows. In section~\ref{sec:Model}, we introduce the $SU(4)/Sp(4)$ FCHM. The calculations of the gauge sector, including SM gauge group and $U(1)'$ gauge symmetry, are presented. To study its phenomenology, we specify the transformation between flavor basis and mass basis in section~\ref{sec:Mixing}. The resulting low energy phenomenology is discussed in section~\ref{sec:Pheno}, including the B anomalies and other experimental constraints. Section~\ref{sec:Collider} focuses on the direct searches, which play an important role in constraining a TeV-scale $Z'$ boson. Section~\ref{sec:Discussion} and Section~\ref{sec:Conclusion} contains our discussions and conclusions.


\section{The $SU(4)/Sp(4)$ FCHM}\label{sec:Model}

In fundamental composite Higgs models, additional hyperfermions $\psi$ are added to generate composite Higgs. The hyperfermions are representations of hypercolor group $G_{HC}$, whose coupling becomes strong around the TeV scale. The hyperfermions then form a condensate, which breaks the global symmetry and results in the pNGBs as the Higgs doublet. In this paper, we study the minimal fundamental composite Higgs model based on the global symmetry breaking $SU(4)\to Sp(4)$. The fermionic UV completion of a $SU(4)/Sp(4)$ FCHM only require four Weyl fermions in the fundamental representation of the $SU(2)=Sp(2)$ hypercolor group \cite{Cacciapaglia:2014uja, Cacciapaglia:2020kgq}. The four Weyl fermions transform under $G_{SM}=SU(3)_C\times SU(2)_L\times U(1)_Y$ as
\begin{align}
&\psi_L=(U_L,D_L)=(1,2,0),\nonumber\\
&U_R=(1,1,1/2),\quad D_R=(1,1,-1/2).
\end{align}

Next, we rewrite the two right-handed hyperfermions as $\tilde{U}_L=-i\sigma^2C\bar{U}^T_R$ and $\tilde{D}_L=-i\sigma^2C\bar{D}^T_R$. Since all the four Weyl fermions are according to the same representation of the hypercolor group, we can recast them together as 
\begin{equation}
\psi=(U_L,D_L,\tilde{U}_L,\tilde{D}_L)^T~,
\end{equation}
which has a $SU(4)$ global symmetry (partially gauged). The hypercolor group becomes strongly coupled at the TeV scale, which forms a non-perturbative vacuum and breaks the $SU(4)$ down to $Sp(4)$. In CHMs, the condensate $\langle \psi\psi \rangle \propto \Sigma_0$ is chosen such that electroweak symmetry is preserved. It will be broken after the Higgs interactions and loop-induced potentials are taken into account. However, we will only focus on some key ingredients here and leave the complete analysis to the future.

\subsection{Basics of $SU(4)/Sp(4)$}

To study the $SU(4)/Sp(4)$ symmetry breaking, we can parametrize it by a non-linear sigma model. Consider a sigma field $\Sigma$, which transforms as an anti-symmetric tensor representation $\mathbf{6}$ of $SU(4)$. The transformation can be expressed as $\Sigma \to g\,\Sigma \,g^T$ with $g\in SU(4)$. The scalar field $\Sigma$ has an anti-symmetric VEV $\langle \Sigma\rangle$, where
\begin{equation}
\langle \Sigma\rangle = \Sigma_0=
\begin{pmatrix}
i\sigma_2  &  0 \\
0   &  i\sigma_2 \\
\end{pmatrix}.
\end{equation}
The $\Sigma$ VEV breaks $SU(4)$ down to $Sp(4)$, producing five Nambu-Goldstone bosons.

The 15 $SU(4)$ generators can be divided into the unbroken ones and broken ones with each type satisfying 
\begin{equation}
\begin{cases}
\text{unbroken generators}   &T_a   : T_a\Sigma_0+\Sigma_0T_a^T=0~,\\
\text{broken generators}       &X_a   : X_a\Sigma_0-\Sigma_0X_a^T=0~.
\end{cases}
\end{equation}
The Nambu-Goldstone fields can be written as a matrix with the broken generator:
\begin{equation}
\xi(x)\equiv e^{\frac{i\pi_a(x)X_a}{2f}}. 
\end{equation}
Under $SU(4)$, the $\xi$ field transforms as $\xi \to g\, \xi \,h^{\dagger}$ where $g \in SU(4)$ and $h \in Sp(4)$. The relation between $\xi$ and $\Sigma$ field is given by
\begin{equation}
\Sigma(x)= \xi\, \Sigma_0\,\xi^T=e^{\frac{i\pi_a(x)X_a}{f}}\Sigma_0~.
\end{equation}

The broken generators and the corresponding fields in the matrix can be organized as follows:
\begin{align}
i\pi_aX_a=&
\begin{pmatrix}
{ia}\,\mathbb{I} & \sqrt{2}\left(\tilde{H}H\right) \\  
-\sqrt{2}\left(\tilde{H}H\right)^\dagger & -ia\,\mathbb{I} \\  
\end{pmatrix}
\label{Goldstone}
\end{align}
In this matrix, there are five independent fields. The four of them form the Higgs (complex) doublet $H$. Besides, there is one more singlet $a$, which will turn out to be the longitudinal part of the $Z'$ boson. By these matrices, we can construct the low energy effective Lagrangian for these pNGB fields.

\subsection{The SM gauge sector}

The SM electroweak gauge group $SU(2)_W\times U(1)_Y$ is embedded in $SU(4)\times U(1)_X$ with generators given by
\begin{equation}
SU(2)_W: \frac{1}{2}
\begin{pmatrix}
\sigma^a   &  0  \\
0   &  0  \\
\end{pmatrix},\quad
U(1)_Y: \frac{1}{2}
\begin{pmatrix}
0   &  0  &  0   &  0  \\
0   &  0  &  0   &  0  \\
0   &  0  &  -1  &  0  \\
0   &  0  &  0   &  1  \\
\end{pmatrix} + X \mathbf{I}~.
\end{equation}
The extra $U(1)_X$ factor accounts for the different hypercharges of the fermion representations but is not relevant for the bosonic fields. These generators belong to $Sp(4)\times U(1)_X$ and are not broken by $\Sigma_0$.
Using the $\Sigma$ field, the Lagrangian for kinetic terms of Higgs boson comes from
\begin{equation}
\mathcal{L}_h=\frac{f^2}{8}\text{tr}\left[(D_{\mu}\Sigma)(D^\mu \Sigma)^\dagger\right]+\cdots ,
\label{Lagrangian}
\end{equation}
where $D_{\mu}$ is the electroweak covariant derivative. Expanding this, we get
\begin{equation}
\mathcal{L}_h=\frac{1}{2}(\partial _\mu h)^2+\frac{f^2}{8}g_W^2 \,\text{sin}^2\left(\frac{h}{f}\right) \left[2W^+_\mu W^{-\mu}+\frac{Z_\mu Z^\mu}{\text{cos}\,\theta_W}\right].
\end{equation}
The non-linear behavior of the Higgs boson in the CHM is apparent from the dependence of trigonometric functions. When $h$ obtains a nonzero VEV $\langle h\rangle=V$, the $W$ boson acquires a mass of
\begin{equation}
m_W^2=\frac{f^2}{4}g_W^2 \,\text{sin}^2\left(\frac{V}{f}\right)=\frac{1}{4}g_W^2v^2,
\end{equation}
where $v\equiv f\,\text{sin}(V/f)\approx V$.
The non-linearity of the CHM is parametrized by
\begin{equation}
\xi\equiv \frac{v^2}{f^2}= \sin^2\left(\frac{V}{f}\right)~.
\end{equation}

The Higgs boson couplings to SM fields in the $SU(4)/Sp(4)$ CHM are modified by  the non-linear effect due to the pNGB nature of the Higgs boson. For example, the deviation of the Higgs coupling to vector bosons is parameterized by
\begin{equation}
\kappa_V\equiv \frac{g_{hVV}}{g^{SM}_{hVV}}=
\text{cos}\left(\frac{V}{f}\right)=\sqrt{1-\xi}\approx 1-\frac{\xi}{2}~.
\end{equation}
To decide the bound on the parameter $\xi$, we also need to determine the Yukawa coupling in the model, which is beyond the scope of the present work. The most conservative bound requires $\xi \lesssim 0.06$ \cite{Sanz:2017tco, ATLAS:2019nkf}, which implies the symmetry breaking scale $f \gtrsim 1$ TeV.

\subsection{$U(1)'$ gauge symmetry}\label{sec:U(1)}

Besides the SM gauge symmetry, we also gauge the $U(1)'$ subgroup of $SU(4)$ with the generator given by
\begin{equation}
U(1)': ~Q_{HC}
\begin{pmatrix}
\mathbb{I}   &  0    \\
0   &  -\mathbb{I}   \\
\end{pmatrix}.
\end{equation}
The $U(1)'$ behaves like the lepton number of hyperfermions, where a hyperfermion carry charge $Q_{HC}$ and an anti-hyperfermion carry charge $-Q_{HC}$. To explain the neutral current B anomalies without violating the experimental constraints, we assume SM fermions (but only the third generation) also carry a nonzero, universal charge, which is set to 1 for simplicity as mentioned in eq. \eqref{int}. To make the $U(1)'$ gauge symmetry anomaly-free, we need to take $Q_{HC}=-2$ in the minimal FCHM. Now the $U(1)'$ gauge symmetry becomes the difference between the third generation SM number and the hyperfermion number, or written as $SM_3-HF$, which is like the hyper version of anomaly-free $B-L$ symmetry.

When $SU(4)$ global symmetry is broken down by the $\Sigma$ VEV to $Sp(4)$ at the symmetry breaking scale, the $U(1)'$ subgroup is also broken down. It results in a massive $Z'$ gauge boson with
\begin{equation}
M_{Z'} =g_{Z'}\left(2\,|Q_{HC}|f\right)\equiv g_{Z'}f',
\end{equation}
where we define the scale 
\begin{equation}
f'\equiv2\,|Q_{HC}|f=4f,
\end{equation}
which is relevant in the study of $Z'$ phenomenology.

To sum up, in this flavorful $SU(4)/Sp(4)$ FCHM, five pNGBs are generated below the compositeness scale. The four of them become the SM Higgs doublet we observed but with non-linear nature, which will be tested in the future Higgs measurements. The 5$^{th}$ one is eaten by the $U(1)'$ gauge boson and results in a heavy $Z'$ boson around the TeV scale. Other model construction issues and phenomenology of $SU(4)/Sp(4)$ CHM have been studied comprehensively in \cite{Cacciapaglia:2014uja, Cacciapaglia:2020kgq}. In the following sections, we will focus on the $Z'$ phenomenology and the connection with the B anomalies.


\section{Specify the mixing matrices for phenomenology}\label{sec:Mixing}

To discuss the phenomenology, we need to first rewrite the $Z'$ interaction terms in eq.~\eqref{int} to cover all generations and separate different chirality as
\begin{align}
\mathcal{L}_{\text{int}}=g_{Z'}Z'_\mu\,(\,\bar{F}_L^f\gamma^\mu  Q_{F_{L}}^fF_L^f+\bar{F}_R^f\gamma^\mu  Q_{F_{R}}^fF_R^f\,),
\end{align}
where $F=(F_1,F_2,F_3)$ includes SM fermions of all the three generations with superscript $f$ for flavor basis. The $3\times 3$ charge matrices in the flavor basis look like
\begin{equation}
Q_{F_{L/R}}^f=
\begin{pmatrix}
0   &  0  &  0    \\
0   &  0  &  0    \\
0   &  0  &  1    \\
\end{pmatrix}.
\end{equation}
However, to study phenomenology, we need to transform them to the mass basis $F_{L/R}^m$ through the mixing matrices as $F^f_{L/R}= U_{F_{L/R}} F^m_{L/R}$. After the transformation, we get
\begin{align}
\mathcal{L}_{\text{int}}=g_{Z'}Z'_\mu\,(\,\bar{F}_L^m\gamma^\mu Q_{F_{L}}^mF_L^m+\bar{F}_R^m\gamma^\mu Q_{F_{R}}^mF_R^m\,),
\end{align}
where the charge matrices becomes
\begin{equation}
Q_{F_{L/R}}^m=U_{F_{L/R}}^\dagger Q_{F_{L/R}}^fU_{F_{L/R}}.
\end{equation}

Therefore, we need to know all the $U_{F_{L/R}}$ to determine the magnitude of each interaction. However, The only information about these unitary transformation matrices is the CKM matrix for quarks and PMNS matrix for leptons. The two relations that need to be satisfied are
\begin{equation}
V_{CKM}\equiv U_{u_L}^\dagger U_{d_L}\quad\text{and}\quad
V_{PMNS}\equiv U_{\nu_L}^\dagger U_{e_L},
\end{equation}
which only tells us about the left-handed part with no information about the right-handed part. Even with these two constraints, they only give the difference between two unitary transformations, but not the individual one. Therefore, we need to make some assumptions about the matrices so there won't be too many parameters.

To simplify the analysis, we assume all the $U_{F_{R}}$ are identity matrices. Therefore, for right-handed fermions, only the third generation joins in the interaction with no flavor changing at all. The couplings are the same for all the right-handed fermions it couples to with coupling strength $g_{Z'}$.

For the left-handed side, due to the observation of $V_{CKM}$ and $V_{PMNS}$, there is a guarantee minimal transformation for $U_{F_{L}}$. Because we only care about the transition between the second and third generation down-type quarks and charged leptons, we will only specify the rotation $\theta_{23}$ between the second and third generation of $U_{d_L}$ and $U_{e_L}$ as
\begin{equation}
U_{F_L}=
\begin{pmatrix}
1   &0   &0    \\
0   &\text{cos}~\theta_F   &  \text{sin}~\theta_F \\
0   &-\text{sin}~\theta_F   &  \text{cos}~\theta_F \\
\end{pmatrix}
\end{equation}
where $F=d,\,e$. Keeping only the angle $\theta_{23}$ is a strong assumption but a good example case for phenomenological study because it avoids some of the most stringent flavor constraints from light fermions and leaves a simple parameter space for analysis. Following this assumption, the rest of the matrices are fixed as $U_{u_L}=V_{CKM}^\dagger U_{d_L}$ and $U_{\nu_L}=V_{PMNS}^\dagger U_{e_L}$. Notice that, although they looks similar, the magnitude we expect for the two angles are quite different. For $\theta_d$, we expect it to be CKM-like, i.e. sin $\theta_d\sim \mathcal{O}(0.01)$. However, for $\theta_e$, it could be as large as sin $\theta_e\sim 1$.

We can then calculate the charge matrices as
\begin{equation}
Q_{F_{L}}^m=
\begin{pmatrix}
0   &  0  &  0    \\
0   &   \text{sin}^2~\theta_F  &  -\frac{1}{2} \,\text{sin}~2\theta_F    \\
0   &  -\frac{1}{2} \,\text{sin}~2\theta_F  &   \text{cos}^2~\theta_F    \\
\end{pmatrix},
\end{equation}
where $F=d,\,e$, and write down all the coupling for left-handed fermions. To study the B anomalies, two of them, $g_{sb}$ and $g_{\mu\mu}$, are especially important, so we further define
\begin{align}
g_{sb}\equiv -g_{Z'}\epsilon_{sb} &\quad\text{with}\quad
\epsilon_{sb}=\frac{1}{2}\,\text{sin}\,2\theta_d,\\
\quad g_{\mu\mu}\equiv g_{Z'}\epsilon_{\mu\mu}
&\quad\text{with}\quad\epsilon_{\mu\mu}=\text{sin}^2\,\theta_e.
\end{align}
We will see later that constraints will be put on the three key parameters: the scale $f'$, the mixings $\epsilon_{sb}$, and $\epsilon_{\mu\mu}$.


\section{Low Energy Phenomenology}\label{sec:Pheno}

With the specified mixing matrices, we can then discuss the parameter space allowed to explain the B anomalies. Also, the constraints from other low energy experiments are presented in this section.

\subsection{Neutral Current B Anomalies}

To explain the observed neutral current B anomalies, an additional negative contribution on $b\to s\mu^+\mu^-$ is required. Based on the assumption we make, after integrating out the $Z'$ boson, we can get the operator
\begin{equation}
\Delta\mathcal{L}=\frac{4G_F}{\sqrt{2}}V_{tb}V_{ts}^*\frac{e^2}{16\pi^2}C_{LL}(\bar{s}_L\gamma^\rho b_L)(\bar{\mu}_L\gamma_\rho \mu_L)
\end{equation}
in the low energy effective Lagrangian with coefficient
\begin{equation}
C_{LL}
=\frac{g_{sb}g_{\mu\mu}}{M_{Z'}^2}~(35~\text{TeV})^2
=-\frac{\epsilon_{sb}\epsilon_{\mu\mu}}{f'^2}~(35~\text{TeV})^2.
\label{Banomaly}
\end{equation}

The global fit value for the Wilson coefficient, considering all rare B decays \cite{Altmannshofer:2021qrr}, gives
\begin{equation}
C_{LL}=-0.82\pm 0.14~,
\label{fitting}
\end{equation}
which requires
\begin{equation}
\frac{\epsilon_{sb}\epsilon_{\mu\mu}}{f'^2}=\frac{1}{(39~\text{TeV})^2}
\implies f'\sim \sqrt{\epsilon_{sb}\epsilon_{\mu\mu}}~(39~\text{TeV}).
\label{bsmumu}
\end{equation}

The generic scale with large mixing angles is $f' \sim 40$ TeV. However, as we mentioned, the value $\epsilon_{sb}\sim\mathcal{O}(0.01)$, which will bring it down to the TeV scale.

\subsection{Neutral Meson Mixing}

The measurement of neutral meson mixing put strong constraints on the $Z'$ solution. Based on our specified mixing matrices, which have suppressed mixings between the first two generations, the $B_s-\bar{B}_s$ mixing turns out to be the strongest constraint. The measurement of mixing parameter \cite{HFLAV:2016hnz} compared with SM prediction by recent lattice data  \cite{King:2019lal} gives the bound on the $\bar{s}bZ'$ vertex as
\begin{equation}
\frac{g_{Z'}}{M_{Z'}}\epsilon_{sb} \leq \frac{1}{194~\text{TeV}}\implies
f' \geq \epsilon_{sb}\cdot 194~(\text{TeV}).
\label{Bsmixing}
\end{equation}

Combining with the requirement from eq. \eqref{bsmumu}, we can rewrite the constraint as
\begin{equation}
f' \leq \epsilon_{\mu\mu}\cdot 7.7~(\text{TeV})~.
\label{minmumu}
\end{equation}
The constraint can be understood as that, in the $b\to s\mu^+\mu^-$ process, the $bs$ side, which is constrained by the $B_s-\bar{B}_s$ mixing measurement, should be extremely suppressed. Therefore, the $\mu\mu$ side needs to be large enough to generate the observed B anomalies. We can also find a hierarchy $\epsilon_{\mu\mu}/\epsilon_{sb}\geq 25$, which leads to the bound $\epsilon_{sb}\leq 0.04$, which is consistent with what we expected.

\subsection{Lepton Flavor Violation Decay}

\begin{figure}[t]
\centering
\includegraphics[width=1.0\linewidth]{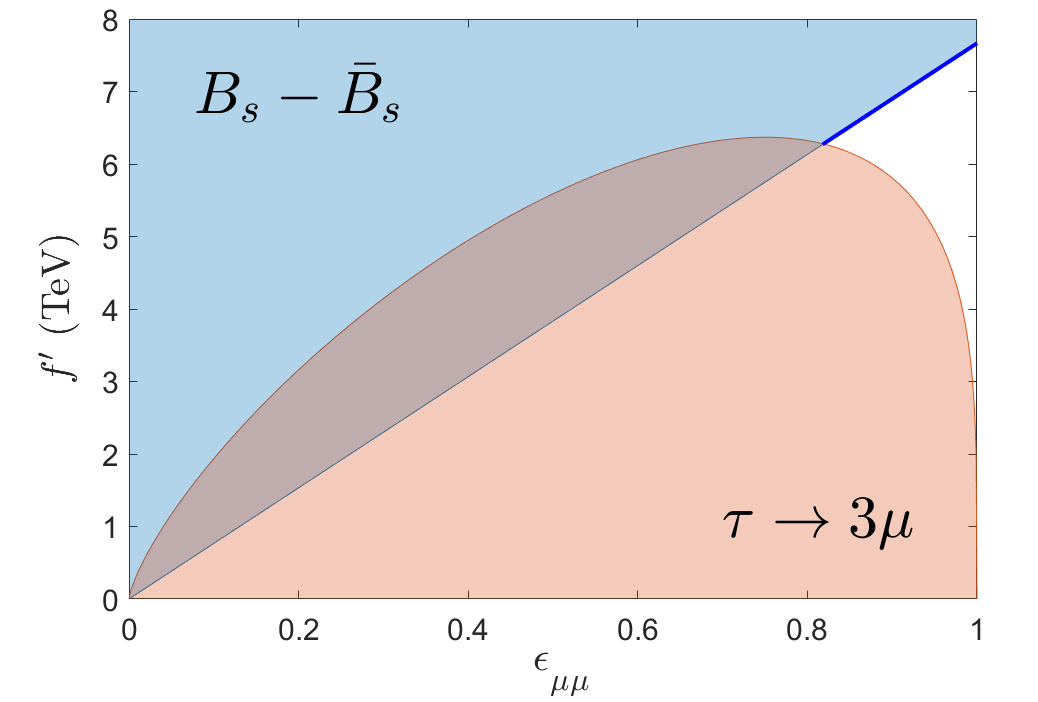}
\caption{The viable parameter space from the experimental constraints. The shaded region is excluded by the corresponding measurements. The bright blue line labels the upper edge of the available parameter space.}
\label{LFV}
\end{figure}

In the lepton sector, there is also a strong constraint from the flavor changing neutral currents (FCNCs). The off-diagonal term in the charge matrix of charged lepton will introduce lepton flavor violation decay, in particular, $\tau\to 3\mu$, from the effective term
\begin{equation}
\mathcal{L}_{LFV}=\frac{g_{Z'}^2}{M_{Z'}^2}s_e^3c_e
(\bar{\tau}_L\gamma^\rho\mu_L)(\bar{\mu}_L\gamma_\rho\mu_L)~,
\end{equation}
where $s_e=\text{sin}\,\theta_e$ and $c_e=\text{cos}\,\theta_e$. The resulting branching ratio can be expressed as
\begin{align}
BR(&\tau\to 3\mu)
=\frac{2m_\tau^5}{1536\pi^3\Gamma_\tau}\left(\frac{g_{Z'}^2}{M_{Z'}^2}s_e^3c_e\right)^2\nonumber\\
&=3.28\times10^{-4}\,\left(\frac{1\text{ TeV}}{f'}\right)^4\epsilon_{\mu\mu}^3(1-\epsilon_{\mu\mu})~.
\end{align}
The value should be $<2.1\times 10^{-8}$ at $90\%$ CL by the measurement \cite{Hayasaka:2010np}. It also puts a strong constraint on the available parameter space. The exclusion plot combining the constraint from $B_s-\bar{B}_s$ mixing on the parameter space $f'$ v.s. $\epsilon_{\mu\mu}$ is shown in Fig.~\ref{LFV}.

The small $\epsilon_{\mu\mu}$ region is excluded, which give a minimal value $\epsilon_{\mu\mu}\geq 0.82$. It implies the angle $\theta_e$ is quite large. The value of $f'$ is bounded from above as shown in eq.~\eqref{minmumu} but not from below as it could be small in the $\epsilon_{\mu\mu}=1$ limit. However, due to the connection with symmetry breaking scale $f \gtrsim 1$ TeV, we are interested in $f' \gtrsim 4$ TeV, which corresponds to the upper region of the parameter space. In this region, the $Z'$ contributions to neutrino trident production \cite{Altmannshofer:2014pba, CCFR:1991lpl} and muon $(g-2)$ \cite{Pospelov:2008zw, Muong-2:2021ojo} are negligible, so we will only focus on the experimental constraints we mention in this section.


\section{Direct $Z'$ Searches}\label{sec:Collider}

The measurements from flavor physics in the last section can only put the constraints on the mixings and the scale $f'=M_{Z'}/g_{Z'}$. The direct searches, on the other hand, can give the lower bound on the mass of $M_{Z'}$ directly. A general $Z'$ collider search has been discussed in \cite{Allanach:2019mfl}. In this section, we will focus on the scenario determined by our model.

\subsection{Decay width and branching ratios}

The partial width of the $Z'$ boson decaying into Weyl fermion pairs $\bar{f_i}f_j$ is 
\begin{equation}
\Gamma_{ij}=\frac{C}{24\pi}g_{ij}^2M_{Z'},
\end{equation}
where $g_{ij}$ is the coupling of $\bar{f_i}f_jZ'$ vertex and $C$ counts the color degree of freedom. In the limit that all $m_f$ are negligible, we get the total relative width as
\begin{equation}
\frac{\Gamma_{Z'}}{M_{Z'}}=\frac{16}{24\pi}g_{Z'}^2\sim 0.2~g_{Z'}^2~.
\end{equation}
The value is important when we try to pick up the bound from the LHC searches.

The dominant decay channels are the diquarks channel of the third generation quarks as
\begin{equation}
Br(t\bar{t})\sim Br(b\bar{b}) \sim 37.5\%.
\end{equation}
Decays to the light quarks and exotic decays like $tc$ and $bs$ are also allowed but strongly suppressed due to the small rotational angles.

The main constraint is expected to come from the clear dilepton channels. Based on the specified mixing matrices we gave, the branching ratios are
\begin{align}
&Br(\tau\tau) \sim 6.25~(1+(1-\epsilon_{\mu\mu})^2)~\%, \\
&Br(\tau\mu) \sim 12.5~\epsilon_{\mu\mu}(1-\epsilon_{\mu\mu})~\%, \\
&Br(\mu\mu) \sim 6.25~\epsilon_{\mu\mu}^2~\%.
\end{align}
We already get $\epsilon_{\mu\mu}\geq 0.82$ from the flavor constraints, which implies $Br(\mu\mu)\geq 4.2 \%$. Therefore, the $\mu\mu$ final state is the most promising channel but also puts the stringent constraint on the $M_{Z'}$.

\subsection{Production cross section}

In the model, the $Z'$ boson only couples to the third generation quarks in the flavor basis. Even after rotating to the mass basis, the couplings to the first and second generation quarks are still suppressed due to the small mixing angles. Therefore, the dominant production come from the process $b\bar{b}\to Z'$. In the following discussion, we will ignore all the other production processes and the small mixing angle $\theta_d$. In this way, the cross section can be written as
\begin{equation}
{\sigma}(b\bar{b}\to Z') \equiv g_{Z'}^2\cdot \sigma_{bb}(M_{Z'})
\end{equation}
where the coupling dependence is taken out. The $\sigma_{bb}$ is determined by the
bottom-quark parton distribution functions \cite{Martin:2009iq, Alwall:2014hca}, which is a function of $M_{Z'}$.

\subsection{The $\mu\mu$ channel search}

\begin{figure}[t]
\centering
\includegraphics[width=1.0\linewidth]{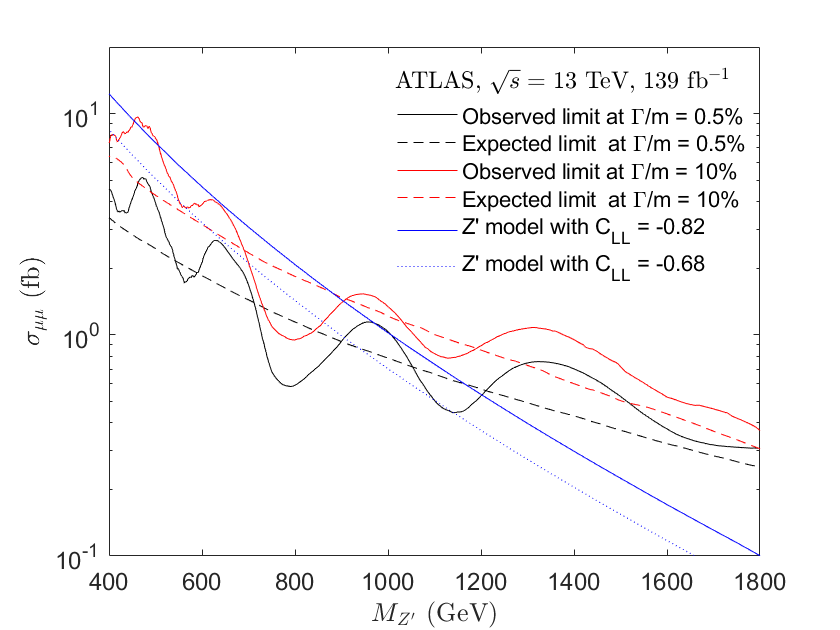}
\caption{Upper limits at 95\% CL on the cross section times branching ratio $\sigma_{\mu\mu}$ as a function of $M_{Z'}$ for 10\% (red) and 0.5\% (black) relative width signals for the dimuon channel. Observed limits are shown as a solid line and expected limits as a dashed line. Also shown are theoretical predictions of the minimal cross section for $Z'$ in the model (blue) assuming $C_{LL}=-0.82$ (solid) and $-0.68$ (dotted).}
\label{sigmamm}
\end{figure}

From the branching ratios and the production cross section we got, we can calculate the cross section for dimuon final state
\begin{equation}
\sigma_{\mu\mu} \equiv \sigma\times Br(\mu\mu) = \frac{1}{16}\,\sigma_{bb}\cdot g_{Z'}^2~\epsilon_{\mu\mu}^2.
\end{equation}
Moreover, from the $B_s-\bar{B}_s$ constraint, we get the lower bound on $\epsilon_{\mu\mu}$ as a function of $f'$ in \eqref{minmumu}, which gives
\begin{equation}
\sigma_{\mu\mu} \geq \frac{1}{16}\,\sigma_{bb}\cdot g_{Z'}^2 \left(\frac{f'}{7.7 \text{ TeV}}\right)^2
=\sigma_{bb} \left(\frac{M_{Z'}}{31 \text{ TeV}}\right)^2.
\end{equation}
The equality holds when $\epsilon_{\mu\mu}= {f'}/{7.7}$ TeV, which corresponds to the blue line in Fig. \ref{LFV}. It gives the minimal cross section as a function of $M_{Z'}$ that allows us to compare with the experimental results. The current best search comes from the ATLAS \cite{ATLAS:2019erb} with an integrated luminosity of 139 fb$^{-1}$. The result is shown in Fig.~\ref{sigmamm}.

Notice that, the bound by collider searches depends on the width. In Fig.~\ref{sigmamm}, we show relative width of 10\%(red) and 0.5\%(black). The wider one gives a weaker bound. However, it require a larger $g_{Z'}\sim 0.7$ and thus a smaller $f'\sim 1.7$ TeV, which is excluded as shown in Fig. \ref{LFV}. The bright blue segment in Fig. \ref{LFV} is the available parameter space with the minimal cross section. In this region, the value $f'\sim 7$ TeV, which implies a smaller $g_{Z'}\sim 0.17$. Therefore, we should use the black line with 0.5\% width in the plot, which requires $M_{Z'}\gtrsim 1200$ GeV. If we relax the best-ﬁt value in the eq. \eqref{fitting} to one sigma region, we get a weaker bound as $M_{Z'}\gtrsim 900$ GeV.

\subsection{Other decay channels}

To looks for other decay channels, we need to first set up benchmark points. From the previous discussion, we choose the value $M_{Z'}=1.4$ TeV, which is right above the current bound. For simplicity, we set $\epsilon_{\mu\mu}=1$, which makes $\sigma_{\tau\tau}=\sigma_{\mu\mu}$ and $\sigma_{\tau\mu}=0$. Once we pick up a value for $f'$, other parameters are automatically set. We can then calculate all the cross sections we are interested in. The results are listed in table \ref{otherdecay}. For a fixed $M_{Z'}$, a larger $f'$ implies a smaller $g_{Z'}$ and thus smaller cross sections. We can check that the $\sigma_{\mu\mu}$ for these benchmark points are still below the bound. Other channels, even with a larger cross section, are well below the observed limits but will be tested during the HL-LHC runs.

\begin{table}[t]
\centering
\begin{tabularx}{0.48\textwidth} 
{|>{\centering\arraybackslash}X|>{\centering\arraybackslash}X|>{\centering\arraybackslash}X|>{\centering\arraybackslash}X|>{\centering\arraybackslash}X|}
\hline
 $f'$(TeV)  & $g_{Z'}$  & $\sigma_{tot}$(fb)  
& $\sigma_{tt/bb}$(fb)  & $\sigma_{\tau\tau/\mu\mu}$(fb)	 \\	\hline	

5.0 	&	0.28 	&	11.21 &							4.20 	&	0.70 			\\ \hline
6.0 	&	0.23 	&	7.79 	&							2.92 	&	0.49 			\\ \hline
7.0 	&	0.20 	&	5.72 	&							2.15 	&	0.36 			\\ \hline

\end{tabularx}
\caption{The cross sections for each decay channel based on $M_{Z'}= 1.4$ TeV with different choice of $f'$.\label{otherdecay}}
\end{table}

We only show the flavor conserving final states so far, but the $Z'$ boson can also have flavor violating decays. However, their cross sections are already constrained by the absence of FCNCs. In the quark sector, the mixings are strongly constrained and thus the branching ratios for these decays are suppressed. However, in the lepton sector, a larger mixing is allowed and the search for flavor violating decays like $Z'\to \mu\tau$ might be viable.

Although other channels are unlikely to be the discovery channel, once the $Z'$ boson is discovered, the next thing to do will be to look for the same resonance in other channels. Through the searches, we can decide the partial widths and figure out the couplings of the $Z'$ boson to other fields. The structure of couplings can help us distinguish between different $Z'$ models. For example, the $Z'$ boson in our model couples universally to all the third generation SM fermions in the flavor basis. Even considering the transformation to the mass basis, it still has a unique partial width ratio
\begin{equation}
\Gamma_{tt}:\Gamma_{bb}:\Gamma_{\ell\ell}:\Gamma_{\nu\nu}\sim 3:3:1:1,
\end{equation}
where $\Gamma_{\ell\ell}$ is the sum of all the charged lepton partial widths. The measurement will allow us to probe the nature of the $Z'$ boson and the underlying  $U(1)'$ symmetry.

\section{Discussions}\label{sec:Discussion}

\begin{figure}[t]
\centering
\includegraphics[width=1.0\linewidth]{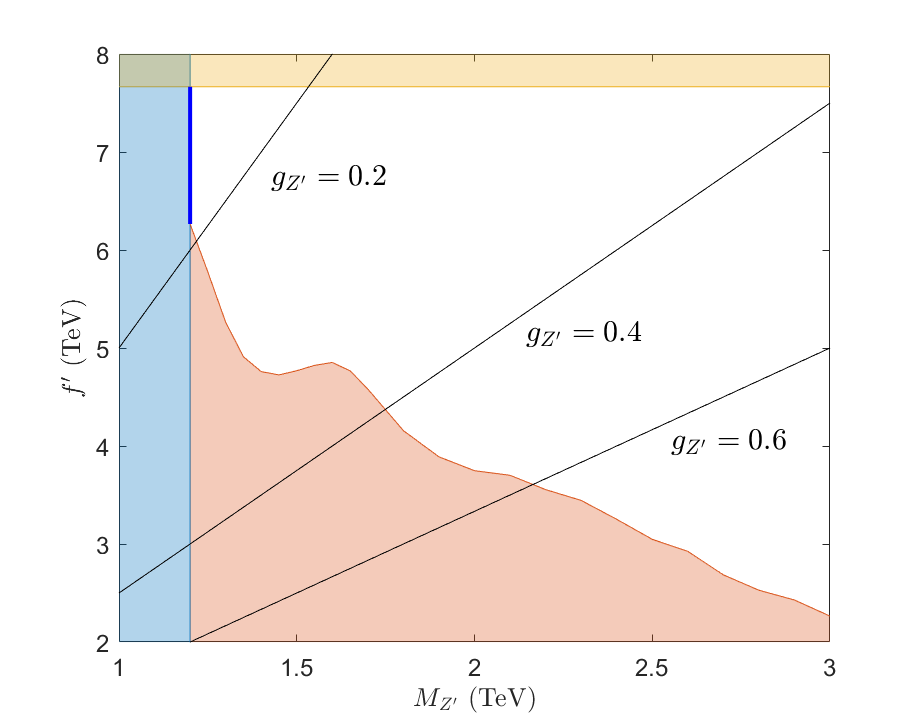}
\caption{Constraints on $f'$ v.s. $M_{Z'}$ plot for $M_{Z'}$ below $3$ TeV. The white region is currently allowed, where $\epsilon_{\mu\mu}$ and $\epsilon_{sb}$ are chosen to satisfy \eqref{Banomaly} from the requirement of the B anomalies. The shaded regions are excluded by the corresponding constraints from Fig.~\ref{LFV} combining with the direct searches, where we use the ATLAS 139 fb$^{-1}$ dimuon searches. The three straight lines represent different values of $g_{Z'}$.}
\label{fvsMz}
\end{figure}

In this study, we are interested in the value of $f'$, which is related to the breaking scale $f$, and the bound on $M_{Z'}$, which is important for the collider searches. 
In the last section, we found that a certain straight line (such as the blue line) in Fig.~\ref{LFV} corresponding to a predicted cross section $\sigma_{\mu\mu}(f'_0)$, which is given by
\begin{equation}
\text{Line: }\epsilon_{\mu\mu}=\frac{f'}{f'_0}\implies
\sigma_{\mu\mu}(f'_0) =\sigma_{bb} \left(\frac{M_{Z'}}{4\times f'_0}\right)^2,
\end{equation}
where $f'_0$ represents the slope of the line, e.g. for the blue line in Fig.~\ref{LFV}, $f'_0=7.7$ TeV. Using this relation, we can calculate the cross section $\sigma_{\mu\mu}$ for each point in the parameter space in Fig.~\ref{LFV} with a certain value of $M_{Z'}$. It allows us to combine ``the constraints in the parameter space in $f'$ v.s. $\epsilon_{\mu\mu}$ plot'' (as shown in Fig.~\ref{LFV}) with ``the direct $\mu\mu$ channel search results from the ATLAS \cite{ATLAS:2019erb}'' into ``the viable parameter space in $f'$ v.s. $M_{Z'}$ plot'' as shown in Fig.~\ref{fvsMz}.

The blue region is excluded by the $B_s-\bar{B}_s$ meson mixing, which gives the lower bound $M_{Z'}\gtrsim 1.2$ TeV. The bright blue line corresponds to the same parameter space as in Fig.~\ref{LFV} with $M_{Z'}\sim 1.2$ TeV. The yellow region, also excluded by the $B_s-\bar{B}_s$ meson mixing, sets the maximum value for $f'$ as shown in eq.~\eqref{minmumu}, which can also be found directly in Fig.~\ref{LFV}. Once the stronger constraint from $B_s-\bar{B}_s$ meson mixing is placed, the yellow line will move downward and the blue line will move rightward. The red region, which is excluded by $\tau\to 3\mu$, restricts the parameter space from below. It places the lower bound on $f'$, which will be pushed upward if the constraint becomes stronger. We can also see the data fluctuations in dimuon search become the fluctuations on the red curve. The strength of the coupling $g_{Z'}$ with three different values is also labeled as the black straight line in the plot.

There are two regions worth noticed in the plot: (1) The region with the light $Z'$ that corresponds to a small $g_{Z'}$ but a large $f'$ region, i.e. $(g_{Z'},f')\sim{(0.2,7 \text{ TeV})}$. (2) For a natural CHM without a large fine-tuning, a smaller $f$ (and thus $f'=4f$) is preferred, which corresponds to a larger $g_{Z'}$ region, such as $(g_{Z'},f')\sim{(0.5,4 \text{ TeV})}$ with a heavier $Z'$. Both regions are around the boundary. The direct searches will extend both blue and red exclusion regions rightward, so both points we mentioned will be probed soon. The lower bound on $M_{Z'}$ will be pushed to 2 TeV and most of the interesting parameter space will be explored during the HL-LHC era \cite{ATL-PHYS-PUB-2018-044, CidVidal:2018eel}.


\section{Conclusions}\label{sec:Conclusion}

In this paper, we presented a new $Z'$ solution to the B anomalies, whose scale is related to the symmetry breaking scale of the underlying strong dynamics. We found that the anomaly-free $U(1)'$ symmetry can arise from $SM_3-HF$, the difference between the third generation SM fermion number and the hyperfermion number. This type of $U(1)'$ is naturally broken at the TeV scale in many fundamental composite Higgs models, which allow us to connect it with the hierarchy problem. We constructed a concrete model based on $SU(4)/Sp(4)$ minimal FCHM. The relation
$f'=2\,|Q_{HC}|f=4f$ connects the flavor anomalies scale $f'$ with the symmetry breaking scale $f$ in the FCHM.

The potential for the $Z'$ boson to explain the B anomalies is discussed in detail. Other flavor physics measurements, like neutral meson mixings and lepton flavor violation decays, put constraints on the allowed parameter space as shown in Fig.~\ref{LFV}. The direct searches also give the bound on the mass of $Z'$ as $M_{Z'}\gtrsim1.2$ TeV. The combined constraints on the scale $f'$ v.s. mass $M_{Z'}$ are shown in Fig.~\ref{fvsMz}, which gives a clear picture about how the parameter space will be probed in the future. Some attractive regions are still viable and will be tested during the HL-LHC era.

\acknowledgments

I thank Hsin-Chia Cheng for many useful discussions. I am also grateful to Ben Allanach and Wolfgang Altmannshofer for reading the previous version and giving many helpful suggestions. This work is supported by the Department of Energy Grant number DE-SC-0009999.

\bibliography{Flavon_Ref}

\begin{thebibliography}{79}%
\makeatletter
\providecommand \@ifxundefined [1]{%
 \@ifx{#1\undefined}
}%
\providecommand \@ifnum [1]{%
 \ifnum #1\expandafter \@firstoftwo
 \else \expandafter \@secondoftwo
 \fi
}%
\providecommand \@ifx [1]{%
 \ifx #1\expandafter \@firstoftwo
 \else \expandafter \@secondoftwo
 \fi
}%
\providecommand \natexlab [1]{#1}%
\providecommand \enquote  [1]{``#1''}%
\providecommand \bibnamefont  [1]{#1}%
\providecommand \bibfnamefont [1]{#1}%
\providecommand \citenamefont [1]{#1}%
\providecommand \href@noop [0]{\@secondoftwo}%
\providecommand \href [0]{\begingroup \@sanitize@url \@href}%
\providecommand \@href[1]{\@@startlink{#1}\@@href}%
\providecommand \@@href[1]{\endgroup#1\@@endlink}%
\providecommand \@sanitize@url [0]{\catcode `\\12\catcode `\$12\catcode
  `\&12\catcode `\#12\catcode `\^12\catcode `\_12\catcode `\%12\relax}%
\providecommand \@@startlink[1]{}%
\providecommand \@@endlink[0]{}%
\providecommand \url  [0]{\begingroup\@sanitize@url \@url }%
\providecommand \@url [1]{\endgroup\@href {#1}{\urlprefix }}%
\providecommand \urlprefix  [0]{URL }%
\providecommand \Eprint [0]{\href }%
\providecommand \doibase [0]{https://doi.org/}%
\providecommand \selectlanguage [0]{\@gobble}%
\providecommand \bibinfo  [0]{\@secondoftwo}%
\providecommand \bibfield  [0]{\@secondoftwo}%
\providecommand \translation [1]{[#1]}%
\providecommand \BibitemOpen [0]{}%
\providecommand \bibitemStop [0]{}%
\providecommand \bibitemNoStop [0]{.\EOS\space}%
\providecommand \EOS [0]{\spacefactor3000\relax}%
\providecommand \BibitemShut  [1]{\csname bibitem#1\endcsname}%
\let\auto@bib@innerbib\@empty
\bibitem [{\citenamefont {Chatrchyan}\ \emph {et~al.}(2012)\citenamefont
  {Chatrchyan} \emph {et~al.}}]{Chatrchyan:2012xdj}%
  \BibitemOpen
  \bibfield  {author} {\bibinfo {author} {\bibfnamefont {S.}~\bibnamefont
  {Chatrchyan}} \emph {et~al.} (\bibinfo {collaboration} {CMS}),\ }\bibfield
  {title} {\bibinfo {title} {{Observation of a new boson at a mass of 125 GeV
  with the CMS experiment at the LHC}},\ }\href
  {https://doi.org/10.1016/j.physletb.2012.08.021} {\bibfield  {journal}
  {\bibinfo  {journal} {Phys. Lett.}\ }\textbf {\bibinfo {volume} {B716}},\
  \bibinfo {pages} {30} (\bibinfo {year} {2012})},\ \Eprint
  {https://arxiv.org/abs/1207.7235} {arXiv:1207.7235 [hep-ex]} \BibitemShut
  {NoStop}%
\bibitem [{\citenamefont {Aad}\ \emph {et~al.}(2012)\citenamefont {Aad} \emph
  {et~al.}}]{Aad:2012tfa}%
  \BibitemOpen
  \bibfield  {author} {\bibinfo {author} {\bibfnamefont {G.}~\bibnamefont
  {Aad}} \emph {et~al.} (\bibinfo {collaboration} {ATLAS}),\ }\bibfield
  {title} {\bibinfo {title} {{Observation of a new particle in the search for
  the Standard Model Higgs boson with the ATLAS detector at the LHC}},\ }\href
  {https://doi.org/10.1016/j.physletb.2012.08.020} {\bibfield  {journal}
  {\bibinfo  {journal} {Phys. Lett.}\ }\textbf {\bibinfo {volume} {B716}},\
  \bibinfo {pages} {1} (\bibinfo {year} {2012})},\ \Eprint
  {https://arxiv.org/abs/1207.7214} {arXiv:1207.7214 [hep-ex]} \BibitemShut
  {NoStop}%
\bibitem [{\citenamefont {Kaplan}\ and\ \citenamefont
  {Georgi}(1984)}]{Kaplan:1983fs}%
  \BibitemOpen
  \bibfield  {author} {\bibinfo {author} {\bibfnamefont {D.~B.}\ \bibnamefont
  {Kaplan}}\ and\ \bibinfo {author} {\bibfnamefont {H.}~\bibnamefont
  {Georgi}},\ }\bibfield  {title} {\bibinfo {title} {{SU(2) x U(1) Breaking by
  Vacuum Misalignment}},\ }\href {https://doi.org/10.1016/0370-2693(84)91177-8}
  {\bibfield  {journal} {\bibinfo  {journal} {Phys. Lett.}\ }\textbf {\bibinfo
  {volume} {136B}},\ \bibinfo {pages} {183} (\bibinfo {year}
  {1984})}\BibitemShut {NoStop}%
\bibitem [{\citenamefont {Kaplan}\ \emph {et~al.}(1984)\citenamefont {Kaplan},
  \citenamefont {Georgi},\ and\ \citenamefont {Dimopoulos}}]{Kaplan:1983sm}%
  \BibitemOpen
  \bibfield  {author} {\bibinfo {author} {\bibfnamefont {D.~B.}\ \bibnamefont
  {Kaplan}}, \bibinfo {author} {\bibfnamefont {H.}~\bibnamefont {Georgi}},\
  and\ \bibinfo {author} {\bibfnamefont {S.}~\bibnamefont {Dimopoulos}},\
  }\bibfield  {title} {\bibinfo {title} {{Composite Higgs Scalars}},\ }\href
  {https://doi.org/10.1016/0370-2693(84)91178-X} {\bibfield  {journal}
  {\bibinfo  {journal} {Phys. Lett.}\ }\textbf {\bibinfo {volume} {136B}},\
  \bibinfo {pages} {187} (\bibinfo {year} {1984})}\BibitemShut {NoStop}%
\bibitem [{\citenamefont {Barnard}\ \emph {et~al.}(2014)\citenamefont
  {Barnard}, \citenamefont {Gherghetta},\ and\ \citenamefont
  {Ray}}]{Barnard:2013zea}%
  \BibitemOpen
  \bibfield  {author} {\bibinfo {author} {\bibfnamefont {J.}~\bibnamefont
  {Barnard}}, \bibinfo {author} {\bibfnamefont {T.}~\bibnamefont
  {Gherghetta}},\ and\ \bibinfo {author} {\bibfnamefont {T.~S.}\ \bibnamefont
  {Ray}},\ }\bibfield  {title} {\bibinfo {title} {{UV descriptions of composite
  Higgs models without elementary scalars}},\ }\href
  {https://doi.org/10.1007/JHEP02(2014)002} {\bibfield  {journal} {\bibinfo
  {journal} {JHEP}\ }\textbf {\bibinfo {volume} {02}},\ \bibinfo {pages}
  {002}},\ \Eprint {https://arxiv.org/abs/1311.6562} {arXiv:1311.6562 [hep-ph]}
  \BibitemShut {NoStop}%
\bibitem [{\citenamefont {Ferretti}\ and\ \citenamefont
  {Karateev}(2014)}]{Ferretti:2013kya}%
  \BibitemOpen
  \bibfield  {author} {\bibinfo {author} {\bibfnamefont {G.}~\bibnamefont
  {Ferretti}}\ and\ \bibinfo {author} {\bibfnamefont {D.}~\bibnamefont
  {Karateev}},\ }\bibfield  {title} {\bibinfo {title} {{Fermionic UV
  completions of Composite Higgs models}},\ }\href
  {https://doi.org/10.1007/JHEP03(2014)077} {\bibfield  {journal} {\bibinfo
  {journal} {JHEP}\ }\textbf {\bibinfo {volume} {03}},\ \bibinfo {pages}
  {077}},\ \Eprint {https://arxiv.org/abs/1312.5330} {arXiv:1312.5330 [hep-ph]}
  \BibitemShut {NoStop}%
\bibitem [{\citenamefont {Cacciapaglia}\ and\ \citenamefont
  {Sannino}(2014)}]{Cacciapaglia:2014uja}%
  \BibitemOpen
  \bibfield  {author} {\bibinfo {author} {\bibfnamefont {G.}~\bibnamefont
  {Cacciapaglia}}\ and\ \bibinfo {author} {\bibfnamefont {F.}~\bibnamefont
  {Sannino}},\ }\bibfield  {title} {\bibinfo {title} {{Fundamental Composite
  (Goldstone) Higgs Dynamics}},\ }\href
  {https://doi.org/10.1007/JHEP04(2014)111} {\bibfield  {journal} {\bibinfo
  {journal} {JHEP}\ }\textbf {\bibinfo {volume} {04}},\ \bibinfo {pages}
  {111}},\ \Eprint {https://arxiv.org/abs/1402.0233} {arXiv:1402.0233 [hep-ph]}
  \BibitemShut {NoStop}%
\bibitem [{\citenamefont {Cacciapaglia}\ \emph {et~al.}(2020)\citenamefont
  {Cacciapaglia}, \citenamefont {Pica},\ and\ \citenamefont
  {Sannino}}]{Cacciapaglia:2020kgq}%
  \BibitemOpen
  \bibfield  {author} {\bibinfo {author} {\bibfnamefont {G.}~\bibnamefont
  {Cacciapaglia}}, \bibinfo {author} {\bibfnamefont {C.}~\bibnamefont {Pica}},\
  and\ \bibinfo {author} {\bibfnamefont {F.}~\bibnamefont {Sannino}},\
  }\bibfield  {title} {\bibinfo {title} {{Fundamental Composite Dynamics: A
  Review}},\ }\href {https://doi.org/10.1016/j.physrep.2020.07.002} {\bibfield
  {journal} {\bibinfo  {journal} {Phys. Rept.}\ }\textbf {\bibinfo {volume}
  {877}},\ \bibinfo {pages} {1} (\bibinfo {year} {2020})},\ \Eprint
  {https://arxiv.org/abs/2002.04914} {arXiv:2002.04914 [hep-ph]} \BibitemShut
  {NoStop}%
\bibitem [{\citenamefont {Katz}\ \emph {et~al.}(2005)\citenamefont {Katz},
  \citenamefont {Nelson},\ and\ \citenamefont {Walker}}]{Katz:2005au}%
  \BibitemOpen
  \bibfield  {author} {\bibinfo {author} {\bibfnamefont {E.}~\bibnamefont
  {Katz}}, \bibinfo {author} {\bibfnamefont {A.~E.}\ \bibnamefont {Nelson}},\
  and\ \bibinfo {author} {\bibfnamefont {D.~G.~E.}\ \bibnamefont {Walker}},\
  }\bibfield  {title} {\bibinfo {title} {{The Intermediate Higgs}},\ }\href
  {https://doi.org/10.1088/1126-6708/2005/08/074} {\bibfield  {journal}
  {\bibinfo  {journal} {JHEP}\ }\textbf {\bibinfo {volume} {08}},\ \bibinfo
  {pages} {074}},\ \Eprint {https://arxiv.org/abs/hep-ph/0504252}
  {arXiv:hep-ph/0504252} \BibitemShut {NoStop}%
\bibitem [{\citenamefont {Gripaios}\ \emph {et~al.}(2009)\citenamefont
  {Gripaios}, \citenamefont {Pomarol}, \citenamefont {Riva},\ and\
  \citenamefont {Serra}}]{Gripaios:2009pe}%
  \BibitemOpen
  \bibfield  {author} {\bibinfo {author} {\bibfnamefont {B.}~\bibnamefont
  {Gripaios}}, \bibinfo {author} {\bibfnamefont {A.}~\bibnamefont {Pomarol}},
  \bibinfo {author} {\bibfnamefont {F.}~\bibnamefont {Riva}},\ and\ \bibinfo
  {author} {\bibfnamefont {J.}~\bibnamefont {Serra}},\ }\bibfield  {title}
  {\bibinfo {title} {{Beyond the Minimal Composite Higgs Model}},\ }\href
  {https://doi.org/10.1088/1126-6708/2009/04/070} {\bibfield  {journal}
  {\bibinfo  {journal} {JHEP}\ }\textbf {\bibinfo {volume} {04}},\ \bibinfo
  {pages} {070}},\ \Eprint {https://arxiv.org/abs/0902.1483} {arXiv:0902.1483
  [hep-ph]} \BibitemShut {NoStop}%
\bibitem [{\citenamefont {Galloway}\ \emph {et~al.}(2010)\citenamefont
  {Galloway}, \citenamefont {Evans}, \citenamefont {Luty},\ and\ \citenamefont
  {Tacchi}}]{Galloway:2010bp}%
  \BibitemOpen
  \bibfield  {author} {\bibinfo {author} {\bibfnamefont {J.}~\bibnamefont
  {Galloway}}, \bibinfo {author} {\bibfnamefont {J.~A.}\ \bibnamefont {Evans}},
  \bibinfo {author} {\bibfnamefont {M.~A.}\ \bibnamefont {Luty}},\ and\
  \bibinfo {author} {\bibfnamefont {R.~A.}\ \bibnamefont {Tacchi}},\ }\bibfield
   {title} {\bibinfo {title} {{Minimal Conformal Technicolor and Precision
  Electroweak Tests}},\ }\href {https://doi.org/10.1007/JHEP10(2010)086}
  {\bibfield  {journal} {\bibinfo  {journal} {JHEP}\ }\textbf {\bibinfo
  {volume} {10}},\ \bibinfo {pages} {086}},\ \Eprint
  {https://arxiv.org/abs/1001.1361} {arXiv:1001.1361 [hep-ph]} \BibitemShut
  {NoStop}%
\bibitem [{\citenamefont {Aaij}\ \emph {et~al.}(2013)\citenamefont {Aaij} \emph
  {et~al.}}]{LHCb:2013ghj}%
  \BibitemOpen
  \bibfield  {author} {\bibinfo {author} {\bibfnamefont {R.}~\bibnamefont
  {Aaij}} \emph {et~al.} (\bibinfo {collaboration} {LHCb}),\ }\bibfield
  {title} {\bibinfo {title} {{Measurement of Form-Factor-Independent
  Observables in the Decay $B^{0} \to K^{*0} \mu^+ \mu^-$}},\ }\href
  {https://doi.org/10.1103/PhysRevLett.111.191801} {\bibfield  {journal}
  {\bibinfo  {journal} {Phys. Rev. Lett.}\ }\textbf {\bibinfo {volume} {111}},\
  \bibinfo {pages} {191801} (\bibinfo {year} {2013})},\ \Eprint
  {https://arxiv.org/abs/1308.1707} {arXiv:1308.1707 [hep-ex]} \BibitemShut
  {NoStop}%
\bibitem [{\citenamefont {Aaij}\ \emph {et~al.}(2014)\citenamefont {Aaij} \emph
  {et~al.}}]{LHCb:2014vgu}%
  \BibitemOpen
  \bibfield  {author} {\bibinfo {author} {\bibfnamefont {R.}~\bibnamefont
  {Aaij}} \emph {et~al.} (\bibinfo {collaboration} {LHCb}),\ }\bibfield
  {title} {\bibinfo {title} {{Test of lepton universality using
  $B^{+}\rightarrow K^{+}\ell^{+}\ell^{-}$ decays}},\ }\href
  {https://doi.org/10.1103/PhysRevLett.113.151601} {\bibfield  {journal}
  {\bibinfo  {journal} {Phys. Rev. Lett.}\ }\textbf {\bibinfo {volume} {113}},\
  \bibinfo {pages} {151601} (\bibinfo {year} {2014})},\ \Eprint
  {https://arxiv.org/abs/1406.6482} {arXiv:1406.6482 [hep-ex]} \BibitemShut
  {NoStop}%
\bibitem [{\citenamefont {Aaij}\ \emph {et~al.}(2016)\citenamefont {Aaij} \emph
  {et~al.}}]{LHCb:2015svh}%
  \BibitemOpen
  \bibfield  {author} {\bibinfo {author} {\bibfnamefont {R.}~\bibnamefont
  {Aaij}} \emph {et~al.} (\bibinfo {collaboration} {LHCb}),\ }\bibfield
  {title} {\bibinfo {title} {{Angular analysis of the $B^{0} \to K^{*0} \mu^{+}
  \mu^{-}$ decay using 3 fb$^{-1}$ of integrated luminosity}},\ }\href
  {https://doi.org/10.1007/JHEP02(2016)104} {\bibfield  {journal} {\bibinfo
  {journal} {JHEP}\ }\textbf {\bibinfo {volume} {02}},\ \bibinfo {pages}
  {104}},\ \Eprint {https://arxiv.org/abs/1512.04442} {arXiv:1512.04442
  [hep-ex]} \BibitemShut {NoStop}%
\bibitem [{\citenamefont {Aaij}\ \emph {et~al.}(2017)\citenamefont {Aaij} \emph
  {et~al.}}]{LHCb:2017avl}%
  \BibitemOpen
  \bibfield  {author} {\bibinfo {author} {\bibfnamefont {R.}~\bibnamefont
  {Aaij}} \emph {et~al.} (\bibinfo {collaboration} {LHCb}),\ }\bibfield
  {title} {\bibinfo {title} {{Test of lepton universality with $B^{0}
  \rightarrow K^{*0}\ell^{+}\ell^{-}$ decays}},\ }\href
  {https://doi.org/10.1007/JHEP08(2017)055} {\bibfield  {journal} {\bibinfo
  {journal} {JHEP}\ }\textbf {\bibinfo {volume} {08}},\ \bibinfo {pages}
  {055}},\ \Eprint {https://arxiv.org/abs/1705.05802} {arXiv:1705.05802
  [hep-ex]} \BibitemShut {NoStop}%
\bibitem [{\citenamefont {Aaij}\ \emph {et~al.}(2019)\citenamefont {Aaij} \emph
  {et~al.}}]{LHCb:2019hip}%
  \BibitemOpen
  \bibfield  {author} {\bibinfo {author} {\bibfnamefont {R.}~\bibnamefont
  {Aaij}} \emph {et~al.} (\bibinfo {collaboration} {LHCb}),\ }\bibfield
  {title} {\bibinfo {title} {{Search for lepton-universality violation in
  $B^+\to K^+\ell^+\ell^-$ decays}},\ }\href
  {https://doi.org/10.1103/PhysRevLett.122.191801} {\bibfield  {journal}
  {\bibinfo  {journal} {Phys. Rev. Lett.}\ }\textbf {\bibinfo {volume} {122}},\
  \bibinfo {pages} {191801} (\bibinfo {year} {2019})},\ \Eprint
  {https://arxiv.org/abs/1903.09252} {arXiv:1903.09252 [hep-ex]} \BibitemShut
  {NoStop}%
\bibitem [{\citenamefont {Aaij}\ \emph {et~al.}(2020)\citenamefont {Aaij} \emph
  {et~al.}}]{LHCb:2020lmf}%
  \BibitemOpen
  \bibfield  {author} {\bibinfo {author} {\bibfnamefont {R.}~\bibnamefont
  {Aaij}} \emph {et~al.} (\bibinfo {collaboration} {LHCb}),\ }\bibfield
  {title} {\bibinfo {title} {{Measurement of $CP$-Averaged Observables in the
  $B^{0}\rightarrow K^{*0}\mu^{+}\mu^{-}$ Decay}},\ }\href
  {https://doi.org/10.1103/PhysRevLett.125.011802} {\bibfield  {journal}
  {\bibinfo  {journal} {Phys. Rev. Lett.}\ }\textbf {\bibinfo {volume} {125}},\
  \bibinfo {pages} {011802} (\bibinfo {year} {2020})},\ \Eprint
  {https://arxiv.org/abs/2003.04831} {arXiv:2003.04831 [hep-ex]} \BibitemShut
  {NoStop}%
\bibitem [{\citenamefont {Aaij}\ \emph {et~al.}(2021)\citenamefont {Aaij} \emph
  {et~al.}}]{LHCb:2021trn}%
  \BibitemOpen
  \bibfield  {author} {\bibinfo {author} {\bibfnamefont {R.}~\bibnamefont
  {Aaij}} \emph {et~al.} (\bibinfo {collaboration} {LHCb}),\ }\bibfield
  {title} {\bibinfo {title} {{Test of lepton universality in beauty-quark
  decays}},\ }\href {https://doi.org/.} {\  (\bibinfo {year} {2021})},\ \Eprint
  {https://arxiv.org/abs/2103.11769} {arXiv:2103.11769 [hep-ex]} \BibitemShut
  {NoStop}%
\bibitem [{\citenamefont {Altmannshofer}\ and\ \citenamefont
  {Stangl}(2021)}]{Altmannshofer:2021qrr}%
  \BibitemOpen
  \bibfield  {author} {\bibinfo {author} {\bibfnamefont {W.}~\bibnamefont
  {Altmannshofer}}\ and\ \bibinfo {author} {\bibfnamefont {P.}~\bibnamefont
  {Stangl}},\ }\bibfield  {title} {\bibinfo {title} {{New Physics in Rare B
  Decays after Moriond 2021}}\ }\href@noop {} {} (\bibinfo {year} {2021}),\
  \Eprint {https://arxiv.org/abs/2103.13370} {arXiv:2103.13370 [hep-ph]}
  \BibitemShut {NoStop}%
\bibitem [{\citenamefont {Cornella}\ \emph {et~al.}(2021)\citenamefont
  {Cornella}, \citenamefont {Faroughy}, \citenamefont {Fuentes-Mart\'\i{}n},
  \citenamefont {Isidori},\ and\ \citenamefont {Neubert}}]{Cornella:2021sby}%
  \BibitemOpen
  \bibfield  {author} {\bibinfo {author} {\bibfnamefont {C.}~\bibnamefont
  {Cornella}}, \bibinfo {author} {\bibfnamefont {D.~A.}\ \bibnamefont
  {Faroughy}}, \bibinfo {author} {\bibfnamefont {J.}~\bibnamefont
  {Fuentes-Mart\'\i{}n}}, \bibinfo {author} {\bibfnamefont {G.}~\bibnamefont
  {Isidori}},\ and\ \bibinfo {author} {\bibfnamefont {M.}~\bibnamefont
  {Neubert}},\ }\bibfield  {title} {\bibinfo {title} {{Reading the footprints
  of the B-meson flavor anomalies}}\ }\href@noop {} {} (\bibinfo {year}
  {2021}),\ \Eprint {https://arxiv.org/abs/2103.16558} {arXiv:2103.16558
  [hep-ph]} \BibitemShut {NoStop}%
\bibitem [{\citenamefont {Geng}\ \emph {et~al.}(2021)\citenamefont {Geng},
  \citenamefont {Grinstein}, \citenamefont {J\"ager}, \citenamefont {Li},
  \citenamefont {Martin~Camalich},\ and\ \citenamefont {Shi}}]{Geng:2021nhg}%
  \BibitemOpen
  \bibfield  {author} {\bibinfo {author} {\bibfnamefont {L.-S.}\ \bibnamefont
  {Geng}}, \bibinfo {author} {\bibfnamefont {B.}~\bibnamefont {Grinstein}},
  \bibinfo {author} {\bibfnamefont {S.}~\bibnamefont {J\"ager}}, \bibinfo
  {author} {\bibfnamefont {S.-Y.}\ \bibnamefont {Li}}, \bibinfo {author}
  {\bibfnamefont {J.}~\bibnamefont {Martin~Camalich}},\ and\ \bibinfo {author}
  {\bibfnamefont {R.-X.}\ \bibnamefont {Shi}},\ }\bibfield  {title} {\bibinfo
  {title} {{Implications of new evidence for lepton-universality violation in
  $b\to s\ell^+\ell^-$ decays}}\ }\href@noop {} {} (\bibinfo {year} {2021}),\
  \Eprint {https://arxiv.org/abs/2103.12738} {arXiv:2103.12738 [hep-ph]}
  \BibitemShut {NoStop}%
\bibitem [{\citenamefont {Alok}\ \emph {et~al.}(2019)\citenamefont {Alok},
  \citenamefont {Dighe}, \citenamefont {Gangal},\ and\ \citenamefont
  {Kumar}}]{Alok:2019ufo}%
  \BibitemOpen
  \bibfield  {author} {\bibinfo {author} {\bibfnamefont {A.~K.}\ \bibnamefont
  {Alok}}, \bibinfo {author} {\bibfnamefont {A.}~\bibnamefont {Dighe}},
  \bibinfo {author} {\bibfnamefont {S.}~\bibnamefont {Gangal}},\ and\ \bibinfo
  {author} {\bibfnamefont {D.}~\bibnamefont {Kumar}},\ }\bibfield  {title}
  {\bibinfo {title} {{Continuing search for new physics in $b \to s \mu \mu$
  decays: two operators at a time}},\ }\href
  {https://doi.org/10.1007/JHEP06(2019)089} {\bibfield  {journal} {\bibinfo
  {journal} {JHEP}\ }\textbf {\bibinfo {volume} {06}},\ \bibinfo {pages}
  {089}},\ \Eprint {https://arxiv.org/abs/1903.09617} {arXiv:1903.09617
  [hep-ph]} \BibitemShut {NoStop}%
\bibitem [{\citenamefont {Alguer\'o}\ \emph {et~al.}(2021)\citenamefont
  {Alguer\'o}, \citenamefont {Capdevila}, \citenamefont {Descotes-Genon},
  \citenamefont {Matias},\ and\ \citenamefont
  {Novoa-Brunet}}]{Alguero:2021anc}%
  \BibitemOpen
  \bibfield  {author} {\bibinfo {author} {\bibfnamefont {M.}~\bibnamefont
  {Alguer\'o}}, \bibinfo {author} {\bibfnamefont {B.}~\bibnamefont
  {Capdevila}}, \bibinfo {author} {\bibfnamefont {S.}~\bibnamefont
  {Descotes-Genon}}, \bibinfo {author} {\bibfnamefont {J.}~\bibnamefont
  {Matias}},\ and\ \bibinfo {author} {\bibfnamefont {M.}~\bibnamefont
  {Novoa-Brunet}},\ }\bibfield  {title} {\bibinfo {title} {{$\boldsymbol{b\to
  s\ell\ell}$ global fits after Moriond 2021 results}},\ }in\ \href@noop {}
  {\emph {\bibinfo {booktitle} {{55th Rencontres de Moriond on QCD and High
  Energy Interactions}}}}\ (\bibinfo {year} {2021})\ \Eprint
  {https://arxiv.org/abs/2104.08921} {arXiv:2104.08921 [hep-ph]} \BibitemShut
  {NoStop}%
\bibitem [{\citenamefont {Carvunis}\ \emph {et~al.}(2021)\citenamefont
  {Carvunis}, \citenamefont {Dettori}, \citenamefont {Gangal}, \citenamefont
  {Guadagnoli},\ and\ \citenamefont {Normand}}]{Carvunis:2021jga}%
  \BibitemOpen
  \bibfield  {author} {\bibinfo {author} {\bibfnamefont {A.}~\bibnamefont
  {Carvunis}}, \bibinfo {author} {\bibfnamefont {F.}~\bibnamefont {Dettori}},
  \bibinfo {author} {\bibfnamefont {S.}~\bibnamefont {Gangal}}, \bibinfo
  {author} {\bibfnamefont {D.}~\bibnamefont {Guadagnoli}},\ and\ \bibinfo
  {author} {\bibfnamefont {C.}~\bibnamefont {Normand}},\ }\bibfield  {title}
  {\bibinfo {title} {{On the effective lifetime of $B_s \to \mu \mu \gamma$}},\
  }\href@noop {} {\bibfield  {journal} {\bibinfo  {journal} {.}\ } (\bibinfo
  {year} {2021})},\ \Eprint {https://arxiv.org/abs/2102.13390}
  {arXiv:2102.13390 [hep-ph]} \BibitemShut {NoStop}%
\bibitem [{\citenamefont {Altmannshofer}\ \emph
  {et~al.}(2014{\natexlab{a}})\citenamefont {Altmannshofer}, \citenamefont
  {Gori}, \citenamefont {Pospelov},\ and\ \citenamefont
  {Yavin}}]{Altmannshofer:2014cfa}%
  \BibitemOpen
  \bibfield  {author} {\bibinfo {author} {\bibfnamefont {W.}~\bibnamefont
  {Altmannshofer}}, \bibinfo {author} {\bibfnamefont {S.}~\bibnamefont {Gori}},
  \bibinfo {author} {\bibfnamefont {M.}~\bibnamefont {Pospelov}},\ and\
  \bibinfo {author} {\bibfnamefont {I.}~\bibnamefont {Yavin}},\ }\bibfield
  {title} {\bibinfo {title} {{Quark flavor transitions in $L_\mu-L_\tau$
  models}},\ }\href {https://doi.org/10.1103/PhysRevD.89.095033} {\bibfield
  {journal} {\bibinfo  {journal} {Phys. Rev. D}\ }\textbf {\bibinfo {volume}
  {89}},\ \bibinfo {pages} {095033} (\bibinfo {year} {2014}{\natexlab{a}})},\
  \Eprint {https://arxiv.org/abs/1403.1269} {arXiv:1403.1269 [hep-ph]}
  \BibitemShut {NoStop}%
\bibitem [{\citenamefont {Altmannshofer}\ and\ \citenamefont
  {Yavin}(2015)}]{Altmannshofer:2015mqa}%
  \BibitemOpen
  \bibfield  {author} {\bibinfo {author} {\bibfnamefont {W.}~\bibnamefont
  {Altmannshofer}}\ and\ \bibinfo {author} {\bibfnamefont {I.}~\bibnamefont
  {Yavin}},\ }\bibfield  {title} {\bibinfo {title} {{Predictions for lepton
  flavor universality violation in rare B decays in models with gauged $L_\mu -
  L_\tau$}},\ }\href {https://doi.org/10.1103/PhysRevD.92.075022} {\bibfield
  {journal} {\bibinfo  {journal} {Phys. Rev. D}\ }\textbf {\bibinfo {volume}
  {92}},\ \bibinfo {pages} {075022} (\bibinfo {year} {2015})},\ \Eprint
  {https://arxiv.org/abs/1508.07009} {arXiv:1508.07009 [hep-ph]} \BibitemShut
  {NoStop}%
\bibitem [{\citenamefont {Altmannshofer}\ \emph {et~al.}(2020)\citenamefont
  {Altmannshofer}, \citenamefont {Davighi},\ and\ \citenamefont
  {Nardecchia}}]{Altmannshofer:2019xda}%
  \BibitemOpen
  \bibfield  {author} {\bibinfo {author} {\bibfnamefont {W.}~\bibnamefont
  {Altmannshofer}}, \bibinfo {author} {\bibfnamefont {J.}~\bibnamefont
  {Davighi}},\ and\ \bibinfo {author} {\bibfnamefont {M.}~\bibnamefont
  {Nardecchia}},\ }\bibfield  {title} {\bibinfo {title} {{Gauging the
  accidental symmetries of the standard model, and implications for the flavor
  anomalies}},\ }\href {https://doi.org/10.1103/PhysRevD.101.015004} {\bibfield
   {journal} {\bibinfo  {journal} {Phys. Rev. D}\ }\textbf {\bibinfo {volume}
  {101}},\ \bibinfo {pages} {015004} (\bibinfo {year} {2020})},\ \Eprint
  {https://arxiv.org/abs/1909.02021} {arXiv:1909.02021 [hep-ph]} \BibitemShut
  {NoStop}%
\bibitem [{\citenamefont {Crivellin}\ \emph {et~al.}(2015)\citenamefont
  {Crivellin}, \citenamefont {D'Ambrosio},\ and\ \citenamefont
  {Heeck}}]{Crivellin:2015mga}%
  \BibitemOpen
  \bibfield  {author} {\bibinfo {author} {\bibfnamefont {A.}~\bibnamefont
  {Crivellin}}, \bibinfo {author} {\bibfnamefont {G.}~\bibnamefont
  {D'Ambrosio}},\ and\ \bibinfo {author} {\bibfnamefont {J.}~\bibnamefont
  {Heeck}},\ }\bibfield  {title} {\bibinfo {title} {{Explaining
  $h\to\mu^\pm\tau^\mp$, $B\to K^* \mu^+\mu^-$ and $B\to K \mu^+\mu^-/B\to K
  e^+e^-$ in a two-Higgs-doublet model with gauged $L_\mu-L_\tau$}},\ }\href
  {https://doi.org/10.1103/PhysRevLett.114.151801} {\bibfield  {journal}
  {\bibinfo  {journal} {Phys. Rev. Lett.}\ }\textbf {\bibinfo {volume} {114}},\
  \bibinfo {pages} {151801} (\bibinfo {year} {2015})},\ \Eprint
  {https://arxiv.org/abs/1501.00993} {arXiv:1501.00993 [hep-ph]} \BibitemShut
  {NoStop}%
\bibitem [{\citenamefont {Crivellin}\ \emph {et~al.}(2017)\citenamefont
  {Crivellin}, \citenamefont {Fuentes-Martin}, \citenamefont {Greljo},\ and\
  \citenamefont {Isidori}}]{Crivellin:2016ejn}%
  \BibitemOpen
  \bibfield  {author} {\bibinfo {author} {\bibfnamefont {A.}~\bibnamefont
  {Crivellin}}, \bibinfo {author} {\bibfnamefont {J.}~\bibnamefont
  {Fuentes-Martin}}, \bibinfo {author} {\bibfnamefont {A.}~\bibnamefont
  {Greljo}},\ and\ \bibinfo {author} {\bibfnamefont {G.}~\bibnamefont
  {Isidori}},\ }\bibfield  {title} {\bibinfo {title} {{Lepton Flavor
  Non-Universality in B decays from Dynamical Yukawas}},\ }\href
  {https://doi.org/10.1016/j.physletb.2016.12.057} {\bibfield  {journal}
  {\bibinfo  {journal} {Phys. Lett. B}\ }\textbf {\bibinfo {volume} {766}},\
  \bibinfo {pages} {77} (\bibinfo {year} {2017})},\ \Eprint
  {https://arxiv.org/abs/1611.02703} {arXiv:1611.02703 [hep-ph]} \BibitemShut
  {NoStop}%
\bibitem [{\citenamefont {Alonso}\ \emph
  {et~al.}(2017{\natexlab{a}})\citenamefont {Alonso}, \citenamefont {Cox},
  \citenamefont {Han},\ and\ \citenamefont {Yanagida}}]{Alonso:2017bff}%
  \BibitemOpen
  \bibfield  {author} {\bibinfo {author} {\bibfnamefont {R.}~\bibnamefont
  {Alonso}}, \bibinfo {author} {\bibfnamefont {P.}~\bibnamefont {Cox}},
  \bibinfo {author} {\bibfnamefont {C.}~\bibnamefont {Han}},\ and\ \bibinfo
  {author} {\bibfnamefont {T.~T.}\ \bibnamefont {Yanagida}},\ }\bibfield
  {title} {\bibinfo {title} {{Anomaly-free local horizontal symmetry and
  anomaly-full rare B-decays}},\ }\href
  {https://doi.org/10.1103/PhysRevD.96.071701} {\bibfield  {journal} {\bibinfo
  {journal} {Phys. Rev. D}\ }\textbf {\bibinfo {volume} {96}},\ \bibinfo
  {pages} {071701} (\bibinfo {year} {2017}{\natexlab{a}})},\ \Eprint
  {https://arxiv.org/abs/1704.08158} {arXiv:1704.08158 [hep-ph]} \BibitemShut
  {NoStop}%
\bibitem [{\citenamefont {Alonso}\ \emph
  {et~al.}(2017{\natexlab{b}})\citenamefont {Alonso}, \citenamefont {Cox},
  \citenamefont {Han},\ and\ \citenamefont {Yanagida}}]{Alonso:2017uky}%
  \BibitemOpen
  \bibfield  {author} {\bibinfo {author} {\bibfnamefont {R.}~\bibnamefont
  {Alonso}}, \bibinfo {author} {\bibfnamefont {P.}~\bibnamefont {Cox}},
  \bibinfo {author} {\bibfnamefont {C.}~\bibnamefont {Han}},\ and\ \bibinfo
  {author} {\bibfnamefont {T.~T.}\ \bibnamefont {Yanagida}},\ }\bibfield
  {title} {\bibinfo {title} {{Flavoured $B-L$ local symmetry and anomalous rare
  $B$ decays}},\ }\href {https://doi.org/10.1016/j.physletb.2017.10.027}
  {\bibfield  {journal} {\bibinfo  {journal} {Phys. Lett. B}\ }\textbf
  {\bibinfo {volume} {774}},\ \bibinfo {pages} {643} (\bibinfo {year}
  {2017}{\natexlab{b}})},\ \Eprint {https://arxiv.org/abs/1705.03858}
  {arXiv:1705.03858 [hep-ph]} \BibitemShut {NoStop}%
\bibitem [{\citenamefont {Bonilla}\ \emph {et~al.}(2018)\citenamefont
  {Bonilla}, \citenamefont {Modak}, \citenamefont {Srivastava},\ and\
  \citenamefont {Valle}}]{Bonilla:2017lsq}%
  \BibitemOpen
  \bibfield  {author} {\bibinfo {author} {\bibfnamefont {C.}~\bibnamefont
  {Bonilla}}, \bibinfo {author} {\bibfnamefont {T.}~\bibnamefont {Modak}},
  \bibinfo {author} {\bibfnamefont {R.}~\bibnamefont {Srivastava}},\ and\
  \bibinfo {author} {\bibfnamefont {J.~W.~F.}\ \bibnamefont {Valle}},\
  }\bibfield  {title} {\bibinfo {title} {{$U(1)_{B_3-3L_\mu}$ gauge symmetry as
  a simple description of $b\to s$ anomalies}},\ }\href
  {https://doi.org/10.1103/PhysRevD.98.095002} {\bibfield  {journal} {\bibinfo
  {journal} {Phys. Rev. D}\ }\textbf {\bibinfo {volume} {98}},\ \bibinfo
  {pages} {095002} (\bibinfo {year} {2018})},\ \Eprint
  {https://arxiv.org/abs/1705.00915} {arXiv:1705.00915 [hep-ph]} \BibitemShut
  {NoStop}%
\bibitem [{\citenamefont {Allanach}(2021)}]{Allanach:2020kss}%
  \BibitemOpen
  \bibfield  {author} {\bibinfo {author} {\bibfnamefont {B.~C.}\ \bibnamefont
  {Allanach}},\ }\bibfield  {title} {\bibinfo {title} {{$U(1)_{B_3-L_2}$
  explanation of the neutral current $B$\ensuremath{-}anomalies}},\ }\href
  {https://doi.org/10.1140/epjc/s10052-021-08855-w} {\bibfield  {journal}
  {\bibinfo  {journal} {Eur. Phys. J. C}\ }\textbf {\bibinfo {volume} {81}},\
  \bibinfo {pages} {56} (\bibinfo {year} {2021})},\ \bibinfo {note} {[Erratum:
  Eur.Phys.J.C 81, 321 (2021)]},\ \Eprint {https://arxiv.org/abs/2009.02197}
  {arXiv:2009.02197 [hep-ph]} \BibitemShut {NoStop}%
\bibitem [{\citenamefont {Allanach}\ and\ \citenamefont
  {Davighi}(2018)}]{Allanach:2018lvl}%
  \BibitemOpen
  \bibfield  {author} {\bibinfo {author} {\bibfnamefont {B.~C.}\ \bibnamefont
  {Allanach}}\ and\ \bibinfo {author} {\bibfnamefont {J.}~\bibnamefont
  {Davighi}},\ }\bibfield  {title} {\bibinfo {title} {{Third family hypercharge
  model for $ {R}_{K^{\left(\ast \right)}} $ and aspects of the fermion mass
  problem}},\ }\href {https://doi.org/10.1007/JHEP12(2018)075} {\bibfield
  {journal} {\bibinfo  {journal} {JHEP}\ }\textbf {\bibinfo {volume} {12}},\
  \bibinfo {pages} {075}},\ \Eprint {https://arxiv.org/abs/1809.01158}
  {arXiv:1809.01158 [hep-ph]} \BibitemShut {NoStop}%
\bibitem [{\citenamefont {Allanach}\ and\ \citenamefont
  {Davighi}(2019)}]{Allanach:2019iiy}%
  \BibitemOpen
  \bibfield  {author} {\bibinfo {author} {\bibfnamefont {B.~C.}\ \bibnamefont
  {Allanach}}\ and\ \bibinfo {author} {\bibfnamefont {J.}~\bibnamefont
  {Davighi}},\ }\bibfield  {title} {\bibinfo {title} {{Naturalising the third
  family hypercharge model for neutral current $B$-anomalies}},\ }\href
  {https://doi.org/10.1140/epjc/s10052-019-7414-z} {\bibfield  {journal}
  {\bibinfo  {journal} {Eur. Phys. J. C}\ }\textbf {\bibinfo {volume} {79}},\
  \bibinfo {pages} {908} (\bibinfo {year} {2019})},\ \Eprint
  {https://arxiv.org/abs/1905.10327} {arXiv:1905.10327 [hep-ph]} \BibitemShut
  {NoStop}%
\bibitem [{\citenamefont {Gauld}\ \emph {et~al.}(2014)\citenamefont {Gauld},
  \citenamefont {Goertz},\ and\ \citenamefont {Haisch}}]{Gauld:2013qba}%
  \BibitemOpen
  \bibfield  {author} {\bibinfo {author} {\bibfnamefont {R.}~\bibnamefont
  {Gauld}}, \bibinfo {author} {\bibfnamefont {F.}~\bibnamefont {Goertz}},\ and\
  \bibinfo {author} {\bibfnamefont {U.}~\bibnamefont {Haisch}},\ }\bibfield
  {title} {\bibinfo {title} {{On minimal $Z'$ explanations of the $B\to
  K^*\mu^+\mu^-$ anomaly}},\ }\href
  {https://doi.org/10.1103/PhysRevD.89.015005} {\bibfield  {journal} {\bibinfo
  {journal} {Phys. Rev. D}\ }\textbf {\bibinfo {volume} {89}},\ \bibinfo
  {pages} {015005} (\bibinfo {year} {2014})},\ \Eprint
  {https://arxiv.org/abs/1308.1959} {arXiv:1308.1959 [hep-ph]} \BibitemShut
  {NoStop}%
\bibitem [{\citenamefont {Buras}\ \emph {et~al.}(2014)\citenamefont {Buras},
  \citenamefont {De~Fazio},\ and\ \citenamefont {Girrbach}}]{Buras:2013dea}%
  \BibitemOpen
  \bibfield  {author} {\bibinfo {author} {\bibfnamefont {A.~J.}\ \bibnamefont
  {Buras}}, \bibinfo {author} {\bibfnamefont {F.}~\bibnamefont {De~Fazio}},\
  and\ \bibinfo {author} {\bibfnamefont {J.}~\bibnamefont {Girrbach}},\
  }\bibfield  {title} {\bibinfo {title} {{331 models facing new $b \to s\mu^+
  \mu^-$ data}},\ }\href {https://doi.org/10.1007/JHEP02(2014)112} {\bibfield
  {journal} {\bibinfo  {journal} {JHEP}\ }\textbf {\bibinfo {volume} {02}},\
  \bibinfo {pages} {112}},\ \Eprint {https://arxiv.org/abs/1311.6729}
  {arXiv:1311.6729 [hep-ph]} \BibitemShut {NoStop}%
\bibitem [{\citenamefont {Buras}\ and\ \citenamefont
  {Girrbach}(2013)}]{Buras:2013qja}%
  \BibitemOpen
  \bibfield  {author} {\bibinfo {author} {\bibfnamefont {A.~J.}\ \bibnamefont
  {Buras}}\ and\ \bibinfo {author} {\bibfnamefont {J.}~\bibnamefont
  {Girrbach}},\ }\bibfield  {title} {\bibinfo {title} {{Left-handed $Z'$ and
  $Z$ FCNC quark couplings facing new $b \to s \mu^+ \mu^-$ data}},\ }\href
  {https://doi.org/10.1007/JHEP12(2013)009} {\bibfield  {journal} {\bibinfo
  {journal} {JHEP}\ }\textbf {\bibinfo {volume} {12}},\ \bibinfo {pages}
  {009}},\ \Eprint {https://arxiv.org/abs/1309.2466} {arXiv:1309.2466 [hep-ph]}
  \BibitemShut {NoStop}%
\bibitem [{\citenamefont {Aristizabal~Sierra}\ \emph
  {et~al.}(2015)\citenamefont {Aristizabal~Sierra}, \citenamefont {Staub},\
  and\ \citenamefont {Vicente}}]{AristizabalSierra:2015vqb}%
  \BibitemOpen
  \bibfield  {author} {\bibinfo {author} {\bibfnamefont {D.}~\bibnamefont
  {Aristizabal~Sierra}}, \bibinfo {author} {\bibfnamefont {F.}~\bibnamefont
  {Staub}},\ and\ \bibinfo {author} {\bibfnamefont {A.}~\bibnamefont
  {Vicente}},\ }\bibfield  {title} {\bibinfo {title} {{Shedding light on the
  $b\to s$ anomalies with a dark sector}},\ }\href
  {https://doi.org/10.1103/PhysRevD.92.015001} {\bibfield  {journal} {\bibinfo
  {journal} {Phys. Rev. D}\ }\textbf {\bibinfo {volume} {92}},\ \bibinfo
  {pages} {015001} (\bibinfo {year} {2015})},\ \Eprint
  {https://arxiv.org/abs/1503.06077} {arXiv:1503.06077 [hep-ph]} \BibitemShut
  {NoStop}%
\bibitem [{\citenamefont {Celis}\ \emph {et~al.}(2015)\citenamefont {Celis},
  \citenamefont {Fuentes-Martin}, \citenamefont {Jung},\ and\ \citenamefont
  {Serodio}}]{Celis:2015ara}%
  \BibitemOpen
  \bibfield  {author} {\bibinfo {author} {\bibfnamefont {A.}~\bibnamefont
  {Celis}}, \bibinfo {author} {\bibfnamefont {J.}~\bibnamefont
  {Fuentes-Martin}}, \bibinfo {author} {\bibfnamefont {M.}~\bibnamefont
  {Jung}},\ and\ \bibinfo {author} {\bibfnamefont {H.}~\bibnamefont
  {Serodio}},\ }\bibfield  {title} {\bibinfo {title} {{Family nonuniversal Z'
  models with protected flavor-changing interactions}},\ }\href
  {https://doi.org/10.1103/PhysRevD.92.015007} {\bibfield  {journal} {\bibinfo
  {journal} {Phys. Rev. D}\ }\textbf {\bibinfo {volume} {92}},\ \bibinfo
  {pages} {015007} (\bibinfo {year} {2015})},\ \Eprint
  {https://arxiv.org/abs/1505.03079} {arXiv:1505.03079 [hep-ph]} \BibitemShut
  {NoStop}%
\bibitem [{\citenamefont {Falkowski}\ \emph {et~al.}(2015)\citenamefont
  {Falkowski}, \citenamefont {Nardecchia},\ and\ \citenamefont
  {Ziegler}}]{Falkowski:2015zwa}%
  \BibitemOpen
  \bibfield  {author} {\bibinfo {author} {\bibfnamefont {A.}~\bibnamefont
  {Falkowski}}, \bibinfo {author} {\bibfnamefont {M.}~\bibnamefont
  {Nardecchia}},\ and\ \bibinfo {author} {\bibfnamefont {R.}~\bibnamefont
  {Ziegler}},\ }\bibfield  {title} {\bibinfo {title} {{Lepton Flavor
  Non-Universality in B-meson Decays from a U(2) Flavor Model}},\ }\href
  {https://doi.org/10.1007/JHEP11(2015)173} {\bibfield  {journal} {\bibinfo
  {journal} {JHEP}\ }\textbf {\bibinfo {volume} {11}},\ \bibinfo {pages}
  {173}},\ \Eprint {https://arxiv.org/abs/1509.01249} {arXiv:1509.01249
  [hep-ph]} \BibitemShut {NoStop}%
\bibitem [{\citenamefont {Chiang}\ \emph {et~al.}(2016)\citenamefont {Chiang},
  \citenamefont {He},\ and\ \citenamefont {Valencia}}]{Chiang:2016qov}%
  \BibitemOpen
  \bibfield  {author} {\bibinfo {author} {\bibfnamefont {C.-W.}\ \bibnamefont
  {Chiang}}, \bibinfo {author} {\bibfnamefont {X.-G.}\ \bibnamefont {He}},\
  and\ \bibinfo {author} {\bibfnamefont {G.}~\bibnamefont {Valencia}},\
  }\bibfield  {title} {\bibinfo {title} {{Z' model for $b\rightarrow
  s{\ell}\bar{\ell}$ flavor anomalies}},\ }\href
  {https://doi.org/10.1103/PhysRevD.93.074003} {\bibfield  {journal} {\bibinfo
  {journal} {Phys. Rev. D}\ }\textbf {\bibinfo {volume} {93}},\ \bibinfo
  {pages} {074003} (\bibinfo {year} {2016})},\ \Eprint
  {https://arxiv.org/abs/1601.07328} {arXiv:1601.07328 [hep-ph]} \BibitemShut
  {NoStop}%
\bibitem [{\citenamefont {Boucenna}\ \emph
  {et~al.}(2016{\natexlab{a}})\citenamefont {Boucenna}, \citenamefont {Celis},
  \citenamefont {Fuentes-Martin}, \citenamefont {Vicente},\ and\ \citenamefont
  {Virto}}]{Boucenna:2016wpr}%
  \BibitemOpen
  \bibfield  {author} {\bibinfo {author} {\bibfnamefont {S.~M.}\ \bibnamefont
  {Boucenna}}, \bibinfo {author} {\bibfnamefont {A.}~\bibnamefont {Celis}},
  \bibinfo {author} {\bibfnamefont {J.}~\bibnamefont {Fuentes-Martin}},
  \bibinfo {author} {\bibfnamefont {A.}~\bibnamefont {Vicente}},\ and\ \bibinfo
  {author} {\bibfnamefont {J.}~\bibnamefont {Virto}},\ }\bibfield  {title}
  {\bibinfo {title} {{Non-abelian gauge extensions for B-decay anomalies}},\
  }\href {https://doi.org/10.1016/j.physletb.2016.06.067} {\bibfield  {journal}
  {\bibinfo  {journal} {Phys. Lett. B}\ }\textbf {\bibinfo {volume} {760}},\
  \bibinfo {pages} {214} (\bibinfo {year} {2016}{\natexlab{a}})},\ \Eprint
  {https://arxiv.org/abs/1604.03088} {arXiv:1604.03088 [hep-ph]} \BibitemShut
  {NoStop}%
\bibitem [{\citenamefont {Boucenna}\ \emph
  {et~al.}(2016{\natexlab{b}})\citenamefont {Boucenna}, \citenamefont {Celis},
  \citenamefont {Fuentes-Martin}, \citenamefont {Vicente},\ and\ \citenamefont
  {Virto}}]{Boucenna:2016qad}%
  \BibitemOpen
  \bibfield  {author} {\bibinfo {author} {\bibfnamefont {S.~M.}\ \bibnamefont
  {Boucenna}}, \bibinfo {author} {\bibfnamefont {A.}~\bibnamefont {Celis}},
  \bibinfo {author} {\bibfnamefont {J.}~\bibnamefont {Fuentes-Martin}},
  \bibinfo {author} {\bibfnamefont {A.}~\bibnamefont {Vicente}},\ and\ \bibinfo
  {author} {\bibfnamefont {J.}~\bibnamefont {Virto}},\ }\bibfield  {title}
  {\bibinfo {title} {{Phenomenology of an $SU(2) \times SU(2) \times U(1)$
  model with lepton-flavour non-universality}},\ }\href
  {https://doi.org/10.1007/JHEP12(2016)059} {\bibfield  {journal} {\bibinfo
  {journal} {JHEP}\ }\textbf {\bibinfo {volume} {12}},\ \bibinfo {pages}
  {059}},\ \Eprint {https://arxiv.org/abs/1608.01349} {arXiv:1608.01349
  [hep-ph]} \BibitemShut {NoStop}%
\bibitem [{\citenamefont {Bhatia}\ \emph {et~al.}(2017)\citenamefont {Bhatia},
  \citenamefont {Chakraborty},\ and\ \citenamefont {Dighe}}]{Bhatia:2017tgo}%
  \BibitemOpen
  \bibfield  {author} {\bibinfo {author} {\bibfnamefont {D.}~\bibnamefont
  {Bhatia}}, \bibinfo {author} {\bibfnamefont {S.}~\bibnamefont
  {Chakraborty}},\ and\ \bibinfo {author} {\bibfnamefont {A.}~\bibnamefont
  {Dighe}},\ }\bibfield  {title} {\bibinfo {title} {{Neutrino mixing and $R_K$
  anomaly in U(1)$_X$ models: a bottom-up approach}},\ }\href
  {https://doi.org/10.1007/JHEP03(2017)117} {\bibfield  {journal} {\bibinfo
  {journal} {JHEP}\ }\textbf {\bibinfo {volume} {03}},\ \bibinfo {pages}
  {117}},\ \Eprint {https://arxiv.org/abs/1701.05825} {arXiv:1701.05825
  [hep-ph]} \BibitemShut {NoStop}%
\bibitem [{\citenamefont {Ko}\ \emph {et~al.}(2017)\citenamefont {Ko},
  \citenamefont {Omura}, \citenamefont {Shigekami},\ and\ \citenamefont
  {Yu}}]{Ko:2017lzd}%
  \BibitemOpen
  \bibfield  {author} {\bibinfo {author} {\bibfnamefont {P.}~\bibnamefont
  {Ko}}, \bibinfo {author} {\bibfnamefont {Y.}~\bibnamefont {Omura}}, \bibinfo
  {author} {\bibfnamefont {Y.}~\bibnamefont {Shigekami}},\ and\ \bibinfo
  {author} {\bibfnamefont {C.}~\bibnamefont {Yu}},\ }\bibfield  {title}
  {\bibinfo {title} {{LHCb anomaly and B physics in flavored Z' models with
  flavored Higgs doublets}},\ }\href
  {https://doi.org/10.1103/PhysRevD.95.115040} {\bibfield  {journal} {\bibinfo
  {journal} {Phys. Rev. D}\ }\textbf {\bibinfo {volume} {95}},\ \bibinfo
  {pages} {115040} (\bibinfo {year} {2017})},\ \Eprint
  {https://arxiv.org/abs/1702.08666} {arXiv:1702.08666 [hep-ph]} \BibitemShut
  {NoStop}%
\bibitem [{\citenamefont {Tang}\ and\ \citenamefont {Wu}(2018)}]{Tang:2017gkz}%
  \BibitemOpen
  \bibfield  {author} {\bibinfo {author} {\bibfnamefont {Y.}~\bibnamefont
  {Tang}}\ and\ \bibinfo {author} {\bibfnamefont {Y.-L.}\ \bibnamefont {Wu}},\
  }\bibfield  {title} {\bibinfo {title} {{Flavor non-universal gauge
  interactions and anomalies in B-meson decays}},\ }\href
  {https://doi.org/10.1088/1674-1137/42/3/033104} {\bibfield  {journal}
  {\bibinfo  {journal} {Chin. Phys. C}\ }\textbf {\bibinfo {volume} {42}},\
  \bibinfo {pages} {033104} (\bibinfo {year} {2018})},\ \bibinfo {note}
  {[Erratum: Chin.Phys.C 44, 069101 (2020)]},\ \Eprint
  {https://arxiv.org/abs/1705.05643} {arXiv:1705.05643 [hep-ph]} \BibitemShut
  {NoStop}%
\bibitem [{\citenamefont {Fuyuto}\ \emph {et~al.}(2018)\citenamefont {Fuyuto},
  \citenamefont {Li},\ and\ \citenamefont {Yu}}]{Fuyuto:2017sys}%
  \BibitemOpen
  \bibfield  {author} {\bibinfo {author} {\bibfnamefont {K.}~\bibnamefont
  {Fuyuto}}, \bibinfo {author} {\bibfnamefont {H.-L.}\ \bibnamefont {Li}},\
  and\ \bibinfo {author} {\bibfnamefont {J.-H.}\ \bibnamefont {Yu}},\
  }\bibfield  {title} {\bibinfo {title} {{Implications of hidden gauged $U(1)$
  model for $B$ anomalies}},\ }\href
  {https://doi.org/10.1103/PhysRevD.97.115003} {\bibfield  {journal} {\bibinfo
  {journal} {Phys. Rev. D}\ }\textbf {\bibinfo {volume} {97}},\ \bibinfo
  {pages} {115003} (\bibinfo {year} {2018})},\ \Eprint
  {https://arxiv.org/abs/1712.06736} {arXiv:1712.06736 [hep-ph]} \BibitemShut
  {NoStop}%
\bibitem [{\citenamefont {Bian}\ \emph {et~al.}(2018)\citenamefont {Bian},
  \citenamefont {Lee},\ and\ \citenamefont {Park}}]{Bian:2017xzg}%
  \BibitemOpen
  \bibfield  {author} {\bibinfo {author} {\bibfnamefont {L.}~\bibnamefont
  {Bian}}, \bibinfo {author} {\bibfnamefont {H.~M.}\ \bibnamefont {Lee}},\ and\
  \bibinfo {author} {\bibfnamefont {C.~B.}\ \bibnamefont {Park}},\ }\bibfield
  {title} {\bibinfo {title} {{$B$-meson anomalies and Higgs physics in flavored
  $U(1)'$ model}},\ }\href {https://doi.org/10.1140/epjc/s10052-018-5777-1}
  {\bibfield  {journal} {\bibinfo  {journal} {Eur. Phys. J. C}\ }\textbf
  {\bibinfo {volume} {78}},\ \bibinfo {pages} {306} (\bibinfo {year} {2018})},\
  \Eprint {https://arxiv.org/abs/1711.08930} {arXiv:1711.08930 [hep-ph]}
  \BibitemShut {NoStop}%
\bibitem [{\citenamefont {King}(2018)}]{King:2018fcg}%
  \BibitemOpen
  \bibfield  {author} {\bibinfo {author} {\bibfnamefont {S.~F.}\ \bibnamefont
  {King}},\ }\bibfield  {title} {\bibinfo {title} {{$ {R}_{K^{\left(*\right)}}
  $ and the origin of Yukawa couplings}},\ }\href
  {https://doi.org/10.1007/JHEP09(2018)069} {\bibfield  {journal} {\bibinfo
  {journal} {JHEP}\ }\textbf {\bibinfo {volume} {09}},\ \bibinfo {pages}
  {069}},\ \Eprint {https://arxiv.org/abs/1806.06780} {arXiv:1806.06780
  [hep-ph]} \BibitemShut {NoStop}%
\bibitem [{\citenamefont {Duan}\ \emph {et~al.}(2019)\citenamefont {Duan},
  \citenamefont {Fan}, \citenamefont {Frank}, \citenamefont {Han},\ and\
  \citenamefont {Yang}}]{Duan:2018akc}%
  \BibitemOpen
  \bibfield  {author} {\bibinfo {author} {\bibfnamefont {G.~H.}\ \bibnamefont
  {Duan}}, \bibinfo {author} {\bibfnamefont {X.}~\bibnamefont {Fan}}, \bibinfo
  {author} {\bibfnamefont {M.}~\bibnamefont {Frank}}, \bibinfo {author}
  {\bibfnamefont {C.}~\bibnamefont {Han}},\ and\ \bibinfo {author}
  {\bibfnamefont {J.~M.}\ \bibnamefont {Yang}},\ }\bibfield  {title} {\bibinfo
  {title} {{A minimal $U(1)^\prime$ extension of MSSM in light of the B decay
  anomaly}},\ }\href {https://doi.org/10.1016/j.physletb.2018.12.005}
  {\bibfield  {journal} {\bibinfo  {journal} {Phys. Lett. B}\ }\textbf
  {\bibinfo {volume} {789}},\ \bibinfo {pages} {54} (\bibinfo {year} {2019})},\
  \Eprint {https://arxiv.org/abs/1808.04116} {arXiv:1808.04116 [hep-ph]}
  \BibitemShut {NoStop}%
\bibitem [{\citenamefont {Calibbi}\ \emph {et~al.}(2020)\citenamefont
  {Calibbi}, \citenamefont {Crivellin}, \citenamefont {Kirk}, \citenamefont
  {Manzari},\ and\ \citenamefont {Vernazza}}]{Calibbi:2019lvs}%
  \BibitemOpen
  \bibfield  {author} {\bibinfo {author} {\bibfnamefont {L.}~\bibnamefont
  {Calibbi}}, \bibinfo {author} {\bibfnamefont {A.}~\bibnamefont {Crivellin}},
  \bibinfo {author} {\bibfnamefont {F.}~\bibnamefont {Kirk}}, \bibinfo {author}
  {\bibfnamefont {C.~A.}\ \bibnamefont {Manzari}},\ and\ \bibinfo {author}
  {\bibfnamefont {L.}~\bibnamefont {Vernazza}},\ }\bibfield  {title} {\bibinfo
  {title} {{$Z^\prime$ models with less-minimal flavour violation}},\ }\href
  {https://doi.org/10.1103/PhysRevD.101.095003} {\bibfield  {journal} {\bibinfo
   {journal} {Phys. Rev. D}\ }\textbf {\bibinfo {volume} {101}},\ \bibinfo
  {pages} {095003} (\bibinfo {year} {2020})},\ \Eprint
  {https://arxiv.org/abs/1910.00014} {arXiv:1910.00014 [hep-ph]} \BibitemShut
  {NoStop}%
\bibitem [{Note1()}]{Note1}%
  \BibitemOpen
  \bibinfo {note} {For our interest, we would like to mention some researches
  aiming at explaining the B anomalies within composite Higgs models. Different
  studies using different features of composite theory to address the problem,
  such as additional composite leptoquarks \cite {Gripaios:2014tna,
  Barbieri:2016las, Marzocca:2018wcf, Fuentes-Martin:2020bnh} or composite
  vector resonances \cite {Niehoff:2015bfa, Niehoff:2015iaa, Carmona:2015ena,
  Carmona:2017fsn, Barbieri:2017tuq, Sannino:2017utc, Chala:2018igk}. However,
  they are all different from this study, where we introduce a new $Z'$
  boson.}\BibitemShut {Stop}%
\bibitem [{\citenamefont {Gripaios}\ \emph {et~al.}(2015)\citenamefont
  {Gripaios}, \citenamefont {Nardecchia},\ and\ \citenamefont
  {Renner}}]{Gripaios:2014tna}%
  \BibitemOpen
  \bibfield  {author} {\bibinfo {author} {\bibfnamefont {B.}~\bibnamefont
  {Gripaios}}, \bibinfo {author} {\bibfnamefont {M.}~\bibnamefont
  {Nardecchia}},\ and\ \bibinfo {author} {\bibfnamefont {S.~A.}\ \bibnamefont
  {Renner}},\ }\bibfield  {title} {\bibinfo {title} {{Composite leptoquarks and
  anomalies in $B$-meson decays}},\ }\href
  {https://doi.org/10.1007/JHEP05(2015)006} {\bibfield  {journal} {\bibinfo
  {journal} {JHEP}\ }\textbf {\bibinfo {volume} {05}},\ \bibinfo {pages}
  {006}},\ \Eprint {https://arxiv.org/abs/1412.1791} {arXiv:1412.1791 [hep-ph]}
  \BibitemShut {NoStop}%
\bibitem [{\citenamefont {Barbieri}\ \emph {et~al.}(2017)\citenamefont
  {Barbieri}, \citenamefont {Murphy},\ and\ \citenamefont
  {Senia}}]{Barbieri:2016las}%
  \BibitemOpen
  \bibfield  {author} {\bibinfo {author} {\bibfnamefont {R.}~\bibnamefont
  {Barbieri}}, \bibinfo {author} {\bibfnamefont {C.~W.}\ \bibnamefont
  {Murphy}},\ and\ \bibinfo {author} {\bibfnamefont {F.}~\bibnamefont
  {Senia}},\ }\bibfield  {title} {\bibinfo {title} {{B-decay Anomalies in a
  Composite Leptoquark Model}},\ }\href
  {https://doi.org/10.1140/epjc/s10052-016-4578-7} {\bibfield  {journal}
  {\bibinfo  {journal} {Eur. Phys. J. C}\ }\textbf {\bibinfo {volume} {77}},\
  \bibinfo {pages} {8} (\bibinfo {year} {2017})},\ \Eprint
  {https://arxiv.org/abs/1611.04930} {arXiv:1611.04930 [hep-ph]} \BibitemShut
  {NoStop}%
\bibitem [{\citenamefont {Blanke}\ and\ \citenamefont
  {Crivellin}(2018)}]{Blanke:2018sro}%
  \BibitemOpen
  \bibfield  {author} {\bibinfo {author} {\bibfnamefont {M.}~\bibnamefont
  {Blanke}}\ and\ \bibinfo {author} {\bibfnamefont {A.}~\bibnamefont
  {Crivellin}},\ }\bibfield  {title} {\bibinfo {title} {{$B$ Meson Anomalies in
  a Pati-Salam Model within the Randall-Sundrum Background}},\ }\href
  {https://doi.org/10.1103/PhysRevLett.121.011801} {\bibfield  {journal}
  {\bibinfo  {journal} {Phys. Rev. Lett.}\ }\textbf {\bibinfo {volume} {121}},\
  \bibinfo {pages} {011801} (\bibinfo {year} {2018})},\ \Eprint
  {https://arxiv.org/abs/1801.07256} {arXiv:1801.07256 [hep-ph]} \BibitemShut
  {NoStop}%
\bibitem [{\citenamefont {Marzocca}(2018)}]{Marzocca:2018wcf}%
  \BibitemOpen
  \bibfield  {author} {\bibinfo {author} {\bibfnamefont {D.}~\bibnamefont
  {Marzocca}},\ }\bibfield  {title} {\bibinfo {title} {{Addressing the
  B-physics anomalies in a fundamental Composite Higgs Model}},\ }\href
  {https://doi.org/10.1007/JHEP07(2018)121} {\bibfield  {journal} {\bibinfo
  {journal} {JHEP}\ }\textbf {\bibinfo {volume} {07}},\ \bibinfo {pages}
  {121}},\ \Eprint {https://arxiv.org/abs/1803.10972} {arXiv:1803.10972
  [hep-ph]} \BibitemShut {NoStop}%
\bibitem [{\citenamefont {Fuentes-Mart\'\i{}n}\ and\ \citenamefont
  {Stangl}(2020)}]{Fuentes-Martin:2020bnh}%
  \BibitemOpen
  \bibfield  {author} {\bibinfo {author} {\bibfnamefont {J.}~\bibnamefont
  {Fuentes-Mart\'\i{}n}}\ and\ \bibinfo {author} {\bibfnamefont
  {P.}~\bibnamefont {Stangl}},\ }\bibfield  {title} {\bibinfo {title}
  {{Third-family quark-lepton unification with a fundamental composite
  Higgs}},\ }\href {https://doi.org/10.1016/j.physletb.2020.135953} {\bibfield
  {journal} {\bibinfo  {journal} {Phys. Lett. B}\ }\textbf {\bibinfo {volume}
  {811}},\ \bibinfo {pages} {135953} (\bibinfo {year} {2020})},\ \Eprint
  {https://arxiv.org/abs/2004.11376} {arXiv:2004.11376 [hep-ph]} \BibitemShut
  {NoStop}%
\bibitem [{\citenamefont {Niehoff}\ \emph {et~al.}(2015)\citenamefont
  {Niehoff}, \citenamefont {Stangl},\ and\ \citenamefont
  {Straub}}]{Niehoff:2015bfa}%
  \BibitemOpen
  \bibfield  {author} {\bibinfo {author} {\bibfnamefont {C.}~\bibnamefont
  {Niehoff}}, \bibinfo {author} {\bibfnamefont {P.}~\bibnamefont {Stangl}},\
  and\ \bibinfo {author} {\bibfnamefont {D.~M.}\ \bibnamefont {Straub}},\
  }\bibfield  {title} {\bibinfo {title} {{Violation of lepton flavour
  universality in composite Higgs models}},\ }\href
  {https://doi.org/10.1016/j.physletb.2015.05.063} {\bibfield  {journal}
  {\bibinfo  {journal} {Phys. Lett. B}\ }\textbf {\bibinfo {volume} {747}},\
  \bibinfo {pages} {182} (\bibinfo {year} {2015})},\ \Eprint
  {https://arxiv.org/abs/1503.03865} {arXiv:1503.03865 [hep-ph]} \BibitemShut
  {NoStop}%
\bibitem [{\citenamefont {Niehoff}\ \emph {et~al.}(2016)\citenamefont
  {Niehoff}, \citenamefont {Stangl},\ and\ \citenamefont
  {Straub}}]{Niehoff:2015iaa}%
  \BibitemOpen
  \bibfield  {author} {\bibinfo {author} {\bibfnamefont {C.}~\bibnamefont
  {Niehoff}}, \bibinfo {author} {\bibfnamefont {P.}~\bibnamefont {Stangl}},\
  and\ \bibinfo {author} {\bibfnamefont {D.~M.}\ \bibnamefont {Straub}},\
  }\bibfield  {title} {\bibinfo {title} {{Direct and indirect signals of
  natural composite Higgs models}},\ }\href
  {https://doi.org/10.1007/JHEP01(2016)119} {\bibfield  {journal} {\bibinfo
  {journal} {JHEP}\ }\textbf {\bibinfo {volume} {01}},\ \bibinfo {pages}
  {119}},\ \Eprint {https://arxiv.org/abs/1508.00569} {arXiv:1508.00569
  [hep-ph]} \BibitemShut {NoStop}%
\bibitem [{\citenamefont {Carmona}\ and\ \citenamefont
  {Goertz}(2016)}]{Carmona:2015ena}%
  \BibitemOpen
  \bibfield  {author} {\bibinfo {author} {\bibfnamefont {A.}~\bibnamefont
  {Carmona}}\ and\ \bibinfo {author} {\bibfnamefont {F.}~\bibnamefont
  {Goertz}},\ }\bibfield  {title} {\bibinfo {title} {{Lepton Flavor and
  Nonuniversality from Minimal Composite Higgs Setups}},\ }\href
  {https://doi.org/10.1103/PhysRevLett.116.251801} {\bibfield  {journal}
  {\bibinfo  {journal} {Phys. Rev. Lett.}\ }\textbf {\bibinfo {volume} {116}},\
  \bibinfo {pages} {251801} (\bibinfo {year} {2016})},\ \Eprint
  {https://arxiv.org/abs/1510.07658} {arXiv:1510.07658 [hep-ph]} \BibitemShut
  {NoStop}%
\bibitem [{\citenamefont {Carmona}\ and\ \citenamefont
  {Goertz}(2018)}]{Carmona:2017fsn}%
  \BibitemOpen
  \bibfield  {author} {\bibinfo {author} {\bibfnamefont {A.}~\bibnamefont
  {Carmona}}\ and\ \bibinfo {author} {\bibfnamefont {F.}~\bibnamefont
  {Goertz}},\ }\bibfield  {title} {\bibinfo {title} {{Recent $B$ physics
  anomalies: a first hint for compositeness?}},\ }\href
  {https://doi.org/10.1140/epjc/s10052-018-6437-1} {\bibfield  {journal}
  {\bibinfo  {journal} {Eur. Phys. J. C}\ }\textbf {\bibinfo {volume} {78}},\
  \bibinfo {pages} {979} (\bibinfo {year} {2018})},\ \Eprint
  {https://arxiv.org/abs/1712.02536} {arXiv:1712.02536 [hep-ph]} \BibitemShut
  {NoStop}%
\bibitem [{\citenamefont {Barbieri}\ and\ \citenamefont
  {Tesi}(2018)}]{Barbieri:2017tuq}%
  \BibitemOpen
  \bibfield  {author} {\bibinfo {author} {\bibfnamefont {R.}~\bibnamefont
  {Barbieri}}\ and\ \bibinfo {author} {\bibfnamefont {A.}~\bibnamefont
  {Tesi}},\ }\bibfield  {title} {\bibinfo {title} {{$B$-decay anomalies in
  Pati-Salam SU(4)}},\ }\href {https://doi.org/10.1140/epjc/s10052-018-5680-9}
  {\bibfield  {journal} {\bibinfo  {journal} {Eur. Phys. J. C}\ }\textbf
  {\bibinfo {volume} {78}},\ \bibinfo {pages} {193} (\bibinfo {year} {2018})},\
  \Eprint {https://arxiv.org/abs/1712.06844} {arXiv:1712.06844 [hep-ph]}
  \BibitemShut {NoStop}%
\bibitem [{\citenamefont {Sannino}\ \emph {et~al.}(2018)\citenamefont
  {Sannino}, \citenamefont {Stangl}, \citenamefont {Straub},\ and\
  \citenamefont {Thomsen}}]{Sannino:2017utc}%
  \BibitemOpen
  \bibfield  {author} {\bibinfo {author} {\bibfnamefont {F.}~\bibnamefont
  {Sannino}}, \bibinfo {author} {\bibfnamefont {P.}~\bibnamefont {Stangl}},
  \bibinfo {author} {\bibfnamefont {D.~M.}\ \bibnamefont {Straub}},\ and\
  \bibinfo {author} {\bibfnamefont {A.~E.}\ \bibnamefont {Thomsen}},\
  }\bibfield  {title} {\bibinfo {title} {{Flavor Physics and Flavor Anomalies
  in Minimal Fundamental Partial Compositeness}},\ }\href
  {https://doi.org/10.1103/PhysRevD.97.115046} {\bibfield  {journal} {\bibinfo
  {journal} {Phys. Rev. D}\ }\textbf {\bibinfo {volume} {97}},\ \bibinfo
  {pages} {115046} (\bibinfo {year} {2018})},\ \Eprint
  {https://arxiv.org/abs/1712.07646} {arXiv:1712.07646 [hep-ph]} \BibitemShut
  {NoStop}%
\bibitem [{\citenamefont {Chala}\ and\ \citenamefont
  {Spannowsky}(2018)}]{Chala:2018igk}%
  \BibitemOpen
  \bibfield  {author} {\bibinfo {author} {\bibfnamefont {M.}~\bibnamefont
  {Chala}}\ and\ \bibinfo {author} {\bibfnamefont {M.}~\bibnamefont
  {Spannowsky}},\ }\bibfield  {title} {\bibinfo {title} {{Behavior of composite
  resonances breaking lepton flavor universality}},\ }\href
  {https://doi.org/10.1103/PhysRevD.98.035010} {\bibfield  {journal} {\bibinfo
  {journal} {Phys. Rev. D}\ }\textbf {\bibinfo {volume} {98}},\ \bibinfo
  {pages} {035010} (\bibinfo {year} {2018})},\ \Eprint
  {https://arxiv.org/abs/1803.02364} {arXiv:1803.02364 [hep-ph]} \BibitemShut
  {NoStop}%
\bibitem [{\citenamefont {Sanz}\ and\ \citenamefont
  {Setford}(2018)}]{Sanz:2017tco}%
  \BibitemOpen
  \bibfield  {author} {\bibinfo {author} {\bibfnamefont {V.}~\bibnamefont
  {Sanz}}\ and\ \bibinfo {author} {\bibfnamefont {J.}~\bibnamefont {Setford}},\
  }\bibfield  {title} {\bibinfo {title} {{Composite Higgs Models after Run
  2}},\ }\href {https://doi.org/10.1155/2018/7168480} {\bibfield  {journal}
  {\bibinfo  {journal} {Adv. High Energy Phys.}\ }\textbf {\bibinfo {volume}
  {2018}},\ \bibinfo {pages} {7168480} (\bibinfo {year} {2018})},\ \Eprint
  {https://arxiv.org/abs/1703.10190} {arXiv:1703.10190 [hep-ph]} \BibitemShut
  {NoStop}%
\bibitem [{\citenamefont {Aad}\ \emph {et~al.}(2020)\citenamefont {Aad} \emph
  {et~al.}}]{ATLAS:2019nkf}%
  \BibitemOpen
  \bibfield  {author} {\bibinfo {author} {\bibfnamefont {G.}~\bibnamefont
  {Aad}} \emph {et~al.} (\bibinfo {collaboration} {ATLAS}),\ }\bibfield
  {title} {\bibinfo {title} {{Combined measurements of Higgs boson production
  and decay using up to $80$ fb$^{-1}$ of proton-proton collision data at
  $\sqrt{s}=$ 13 TeV collected with the ATLAS experiment}},\ }\href
  {https://doi.org/10.1103/PhysRevD.101.012002} {\bibfield  {journal} {\bibinfo
   {journal} {Phys. Rev. D}\ }\textbf {\bibinfo {volume} {101}},\ \bibinfo
  {pages} {012002} (\bibinfo {year} {2020})},\ \Eprint
  {https://arxiv.org/abs/1909.02845} {arXiv:1909.02845 [hep-ex]} \BibitemShut
  {NoStop}%
\bibitem [{\citenamefont {Amhis}\ \emph {et~al.}(2017)\citenamefont {Amhis}
  \emph {et~al.}}]{HFLAV:2016hnz}%
  \BibitemOpen
  \bibfield  {author} {\bibinfo {author} {\bibfnamefont {Y.}~\bibnamefont
  {Amhis}} \emph {et~al.} (\bibinfo {collaboration} {HFLAV}),\ }\bibfield
  {title} {\bibinfo {title} {{Averages of $b$-hadron, $c$-hadron, and
  $\tau$-lepton properties as of summer 2016}},\ }\href
  {https://doi.org/10.1140/epjc/s10052-017-5058-4} {\bibfield  {journal}
  {\bibinfo  {journal} {Eur. Phys. J. C}\ }\textbf {\bibinfo {volume} {77}},\
  \bibinfo {pages} {895} (\bibinfo {year} {2017})},\ \Eprint
  {https://arxiv.org/abs/1612.07233} {arXiv:1612.07233 [hep-ex]} \BibitemShut
  {NoStop}%
\bibitem [{\citenamefont {King}\ \emph {et~al.}(2019)\citenamefont {King},
  \citenamefont {Lenz},\ and\ \citenamefont {Rauh}}]{King:2019lal}%
  \BibitemOpen
  \bibfield  {author} {\bibinfo {author} {\bibfnamefont {D.}~\bibnamefont
  {King}}, \bibinfo {author} {\bibfnamefont {A.}~\bibnamefont {Lenz}},\ and\
  \bibinfo {author} {\bibfnamefont {T.}~\bibnamefont {Rauh}},\ }\bibfield
  {title} {\bibinfo {title} {{B$_{s}$ mixing observables and
  |V$_{td}$/V$_{ts}$| from sum rules}},\ }\href
  {https://doi.org/10.1007/JHEP05(2019)034} {\bibfield  {journal} {\bibinfo
  {journal} {JHEP}\ }\textbf {\bibinfo {volume} {05}},\ \bibinfo {pages}
  {034}},\ \Eprint {https://arxiv.org/abs/1904.00940} {arXiv:1904.00940
  [hep-ph]} \BibitemShut {NoStop}%
\bibitem [{\citenamefont {Hayasaka}\ \emph {et~al.}(2010)\citenamefont
  {Hayasaka} \emph {et~al.}}]{Hayasaka:2010np}%
  \BibitemOpen
  \bibfield  {author} {\bibinfo {author} {\bibfnamefont {K.}~\bibnamefont
  {Hayasaka}} \emph {et~al.},\ }\bibfield  {title} {\bibinfo {title} {{Search
  for Lepton Flavor Violating Tau Decays into Three Leptons with 719 Million
  Produced Tau+Tau- Pairs}},\ }\href
  {https://doi.org/10.1016/j.physletb.2010.03.037} {\bibfield  {journal}
  {\bibinfo  {journal} {Phys. Lett. B}\ }\textbf {\bibinfo {volume} {687}},\
  \bibinfo {pages} {139} (\bibinfo {year} {2010})},\ \Eprint
  {https://arxiv.org/abs/1001.3221} {arXiv:1001.3221 [hep-ex]} \BibitemShut
  {NoStop}%
\bibitem [{\citenamefont {Altmannshofer}\ \emph
  {et~al.}(2014{\natexlab{b}})\citenamefont {Altmannshofer}, \citenamefont
  {Gori}, \citenamefont {Pospelov},\ and\ \citenamefont
  {Yavin}}]{Altmannshofer:2014pba}%
  \BibitemOpen
  \bibfield  {author} {\bibinfo {author} {\bibfnamefont {W.}~\bibnamefont
  {Altmannshofer}}, \bibinfo {author} {\bibfnamefont {S.}~\bibnamefont {Gori}},
  \bibinfo {author} {\bibfnamefont {M.}~\bibnamefont {Pospelov}},\ and\
  \bibinfo {author} {\bibfnamefont {I.}~\bibnamefont {Yavin}},\ }\bibfield
  {title} {\bibinfo {title} {{Neutrino Trident Production: A Powerful Probe of
  New Physics with Neutrino Beams}},\ }\href
  {https://doi.org/10.1103/PhysRevLett.113.091801} {\bibfield  {journal}
  {\bibinfo  {journal} {Phys. Rev. Lett.}\ }\textbf {\bibinfo {volume} {113}},\
  \bibinfo {pages} {091801} (\bibinfo {year} {2014}{\natexlab{b}})},\ \Eprint
  {https://arxiv.org/abs/1406.2332} {arXiv:1406.2332 [hep-ph]} \BibitemShut
  {NoStop}%
\bibitem [{\citenamefont {Mishra}\ \emph {et~al.}(1991)\citenamefont {Mishra}
  \emph {et~al.}}]{CCFR:1991lpl}%
  \BibitemOpen
  \bibfield  {author} {\bibinfo {author} {\bibfnamefont {S.~R.}\ \bibnamefont
  {Mishra}} \emph {et~al.} (\bibinfo {collaboration} {CCFR}),\ }\bibfield
  {title} {\bibinfo {title} {{Neutrino tridents and W Z interference}},\ }\href
  {https://doi.org/10.1103/PhysRevLett.66.3117} {\bibfield  {journal} {\bibinfo
   {journal} {Phys. Rev. Lett.}\ }\textbf {\bibinfo {volume} {66}},\ \bibinfo
  {pages} {3117} (\bibinfo {year} {1991})}\BibitemShut {NoStop}%
\bibitem [{\citenamefont {Pospelov}(2009)}]{Pospelov:2008zw}%
  \BibitemOpen
  \bibfield  {author} {\bibinfo {author} {\bibfnamefont {M.}~\bibnamefont
  {Pospelov}},\ }\bibfield  {title} {\bibinfo {title} {{Secluded U(1) below the
  weak scale}},\ }\href {https://doi.org/10.1103/PhysRevD.80.095002} {\bibfield
   {journal} {\bibinfo  {journal} {Phys. Rev. D}\ }\textbf {\bibinfo {volume}
  {80}},\ \bibinfo {pages} {095002} (\bibinfo {year} {2009})},\ \Eprint
  {https://arxiv.org/abs/0811.1030} {arXiv:0811.1030 [hep-ph]} \BibitemShut
  {NoStop}%
\bibitem [{\citenamefont {Abi}\ \emph {et~al.}(2021)\citenamefont {Abi} \emph
  {et~al.}}]{Muong-2:2021ojo}%
  \BibitemOpen
  \bibfield  {author} {\bibinfo {author} {\bibfnamefont {B.}~\bibnamefont
  {Abi}} \emph {et~al.} (\bibinfo {collaboration} {Muon g-2}),\ }\bibfield
  {title} {\bibinfo {title} {{Measurement of the Positive Muon Anomalous
  Magnetic Moment to 0.46 ppm}},\ }\href
  {https://doi.org/10.1103/PhysRevLett.126.141801} {\bibfield  {journal}
  {\bibinfo  {journal} {Phys. Rev. Lett.}\ }\textbf {\bibinfo {volume} {126}},\
  \bibinfo {pages} {141801} (\bibinfo {year} {2021})},\ \Eprint
  {https://arxiv.org/abs/2104.03281} {arXiv:2104.03281 [hep-ex]} \BibitemShut
  {NoStop}%
\bibitem [{\citenamefont {Allanach}\ \emph {et~al.}(2019)\citenamefont
  {Allanach}, \citenamefont {Butterworth},\ and\ \citenamefont
  {Corbett}}]{Allanach:2019mfl}%
  \BibitemOpen
  \bibfield  {author} {\bibinfo {author} {\bibfnamefont {B.~C.}\ \bibnamefont
  {Allanach}}, \bibinfo {author} {\bibfnamefont {J.~M.}\ \bibnamefont
  {Butterworth}},\ and\ \bibinfo {author} {\bibfnamefont {T.}~\bibnamefont
  {Corbett}},\ }\bibfield  {title} {\bibinfo {title} {{Collider constraints on
  $Z'$ models for neutral current B-anomalies}},\ }\href
  {https://doi.org/10.1007/JHEP08(2019)106} {\bibfield  {journal} {\bibinfo
  {journal} {JHEP}\ }\textbf {\bibinfo {volume} {08}},\ \bibinfo {pages}
  {106}},\ \Eprint {https://arxiv.org/abs/1904.10954} {arXiv:1904.10954
  [hep-ph]} \BibitemShut {NoStop}%
\bibitem [{\citenamefont {Martin}\ \emph {et~al.}(2009)\citenamefont {Martin},
  \citenamefont {Stirling}, \citenamefont {Thorne},\ and\ \citenamefont
  {Watt}}]{Martin:2009iq}%
  \BibitemOpen
  \bibfield  {author} {\bibinfo {author} {\bibfnamefont {A.~D.}\ \bibnamefont
  {Martin}}, \bibinfo {author} {\bibfnamefont {W.~J.}\ \bibnamefont
  {Stirling}}, \bibinfo {author} {\bibfnamefont {R.~S.}\ \bibnamefont
  {Thorne}},\ and\ \bibinfo {author} {\bibfnamefont {G.}~\bibnamefont {Watt}},\
  }\bibfield  {title} {\bibinfo {title} {{Parton distributions for the LHC}},\
  }\href {https://doi.org/10.1140/epjc/s10052-009-1072-5} {\bibfield  {journal}
  {\bibinfo  {journal} {Eur. Phys. J. C}\ }\textbf {\bibinfo {volume} {63}},\
  \bibinfo {pages} {189} (\bibinfo {year} {2009})},\ \Eprint
  {https://arxiv.org/abs/0901.0002} {arXiv:0901.0002 [hep-ph]} \BibitemShut
  {NoStop}%
\bibitem [{\citenamefont {Alwall}\ \emph {et~al.}(2014)\citenamefont {Alwall},
  \citenamefont {Frederix}, \citenamefont {Frixione}, \citenamefont {Hirschi},
  \citenamefont {Maltoni}, \citenamefont {Mattelaer}, \citenamefont {Shao},
  \citenamefont {Stelzer}, \citenamefont {Torrielli},\ and\ \citenamefont
  {Zaro}}]{Alwall:2014hca}%
  \BibitemOpen
  \bibfield  {author} {\bibinfo {author} {\bibfnamefont {J.}~\bibnamefont
  {Alwall}}, \bibinfo {author} {\bibfnamefont {R.}~\bibnamefont {Frederix}},
  \bibinfo {author} {\bibfnamefont {S.}~\bibnamefont {Frixione}}, \bibinfo
  {author} {\bibfnamefont {V.}~\bibnamefont {Hirschi}}, \bibinfo {author}
  {\bibfnamefont {F.}~\bibnamefont {Maltoni}}, \bibinfo {author} {\bibfnamefont
  {O.}~\bibnamefont {Mattelaer}}, \bibinfo {author} {\bibfnamefont {H.~S.}\
  \bibnamefont {Shao}}, \bibinfo {author} {\bibfnamefont {T.}~\bibnamefont
  {Stelzer}}, \bibinfo {author} {\bibfnamefont {P.}~\bibnamefont {Torrielli}},\
  and\ \bibinfo {author} {\bibfnamefont {M.}~\bibnamefont {Zaro}},\ }\bibfield
  {title} {\bibinfo {title} {{The automated computation of tree-level and
  next-to-leading order differential cross sections, and their matching to
  parton shower simulations}},\ }\href
  {https://doi.org/10.1007/JHEP07(2014)079} {\bibfield  {journal} {\bibinfo
  {journal} {JHEP}\ }\textbf {\bibinfo {volume} {07}},\ \bibinfo {pages}
  {079}},\ \Eprint {https://arxiv.org/abs/1405.0301} {arXiv:1405.0301 [hep-ph]}
  \BibitemShut {NoStop}%
\bibitem [{\citenamefont {Aad}\ \emph {et~al.}(2019)\citenamefont {Aad} \emph
  {et~al.}}]{ATLAS:2019erb}%
  \BibitemOpen
  \bibfield  {author} {\bibinfo {author} {\bibfnamefont {G.}~\bibnamefont
  {Aad}} \emph {et~al.} (\bibinfo {collaboration} {ATLAS}),\ }\bibfield
  {title} {\bibinfo {title} {{Search for high-mass dilepton resonances using
  139 fb$^{-1}$ of $pp$ collision data collected at $\sqrt{s}=$13 TeV with the
  ATLAS detector}},\ }\href {https://doi.org/10.1016/j.physletb.2019.07.016}
  {\bibfield  {journal} {\bibinfo  {journal} {Phys. Lett. B}\ }\textbf
  {\bibinfo {volume} {796}},\ \bibinfo {pages} {68} (\bibinfo {year} {2019})},\
  \Eprint {https://arxiv.org/abs/1903.06248} {arXiv:1903.06248 [hep-ex]}
  \BibitemShut {NoStop}%
\bibitem [{ATL(2018)}]{ATL-PHYS-PUB-2018-044}%
  \BibitemOpen
  \href {http://cds.cern.ch/record/2650549} {\emph {\bibinfo {title}
  {{Prospects for searches for heavy $Z^\prime$ and $W^\prime$ bosons in
  fermionic final states with the ATLAS experiment at the HL-LHC}}}},\ \bibinfo
  {type} {Tech. Rep.}\ (\bibinfo  {institution} {CERN},\ \bibinfo {address}
  {Geneva},\ \bibinfo {year} {2018})\BibitemShut {NoStop}%
\bibitem [{\citenamefont {Cid~Vidal}\ \emph {et~al.}(2019)\citenamefont
  {Cid~Vidal} \emph {et~al.}}]{CidVidal:2018eel}%
  \BibitemOpen
  \bibfield  {author} {\bibinfo {author} {\bibfnamefont {X.}~\bibnamefont
  {Cid~Vidal}} \emph {et~al.},\ }\bibfield  {title} {\bibinfo {title} {{Report
  from Working Group 3}: {Beyond the Standard Model physics at the HL-LHC and
  HE-LHC}},\ }\href {https://doi.org/10.23731/CYRM-2019-007.585} {\bibfield
  {journal} {\bibinfo  {journal} {CERN Yellow Rep. Monogr.}\ }\textbf {\bibinfo
  {volume} {7}},\ \bibinfo {pages} {585} (\bibinfo {year} {2019})},\ \Eprint
  {https://arxiv.org/abs/1812.07831} {arXiv:1812.07831 [hep-ph]} \BibitemShut
  {NoStop}%
\end{thebibliography}%

\end{document}